%% file: process-model-application-v4.tex
\documentclass{article}
\usepackage{arxiv}

\usepackage{textcomp}
\usepackage{lscape}
\usepackage{pdflscape}
\usepackage{graphicx}

\usepackage{multibib}

\usepackage[utf8]{inputenc} 
\usepackage[T1]{fontenc}    

\usepackage {hyperref}       

\hypersetup{colorlinks,citecolor=black,linkcolor=black, urlcolor=black}

\usepackage{url}            
\usepackage{booktabs}       
\usepackage{amsfonts}       
\usepackage{nicefrac}       
\usepackage{microtype}      
\usepackage{lipsum}        

\title{Applications of a Novel Knowledge Discovery and Data Mining Process Model for Metabolomics}

\author{
  Ahmed ~BaniMustafa\thanks{ Corresponding author \texttt{banimustafa@gmail.com}. The authors would like to thank the department of computer science at Aberystwyth university for supporting this research} \\
  Department of Computer Science\\
  American University of Madaba\\
  Kings Highway, Madaba, Jordan \\
   \And
  Nigel Hardy \\
  Department of Computer Science\\
  Aberystwyth University, Penglais\\
  Aberystwyth, Ceredigion, UK\\
}

\begin{document}
\maketitle

\begin{abstract}
This work demonstrates the execution of a novel process model for knowledge discovery and data mining for metabolomics (MeKDDaM). It aims to illustrate MeKDDaM process model applicability using four different real-world applications and to highlight its strengths and unique features. The demonstrated applications provide coverage for metabolite profiling, target analysis, and metabolic fingerprinting. The data analysed in these applications were captured by chromatographic separation and mass spectrometry technique (LC-MS), Fourier transform infrared spectroscopy (FT-IR), and nuclear magnetic resonance spectroscopy (NMR) and involve the analysis of plant, animal, and human samples. The process was executed using both data-driven and hypothesis-driven data mining approaches in order to perform various data mining goals and tasks by applying a number of data mining techniques. The applications were selected to achieve a range of analytical goals and research questions and to provide coverage for metabolite profiling, target analysis, and metabolic fingerprinting using datasets that were captured by NMR, LC-MS, and FT-IR using samples of a plant, animal, and human origin. The process was applied using an implementation environment which was created in order to provide a computer-aided realisation of the process model execution.
\end{abstract}

\section{Introduction}

This paper presents a number of metabolomics applications that were used for demonstrating a novel knowledge discovery and data mining process model that was for metabolomics (MeKDDaM), which was designed to provide a justifiable, traceable and reproducible data analysis that achieves the analytical objectives of metabolomics studies and suited to satisfy the nature and requirement of their data. The process model pays attention to the practical aspects of data mining: project management, human interaction, standards support and quality assurance. It also offers an improvement to existing data mining process models regarding its layout structure and scientific orientation based on the principles of scientific methodology, process engineering, software engineering and machine learning. The process provides a number of desirable features seen in many metabolomics investigations: \textbf{data exploration}, knowledge representation, visualization and automation. The applications presented in this work were carried out in order to evaluate and demonstrate the applicability of this process model using four real-life metabolomics applications.

\textbf{Metabolomics}~is defined as: \textgravedbl \emph{the study of all low molecule weight chemicals (metabolites) which are involved in metabolism, either as an end product or as necessary chemicals for metabolism}\textacutedbl \cite{Dun05,Det04,Mal04}, while \textbf{Knowledge discovery} is defined as:\textgravedbl \emph{ the non-trivial process of identifying valid, novel, potentially useful, and ultimately understandable patterns in data}\textacutedbl. Data mining is used by many as a synonym for knowledge discovery, while some considered data mining only a stage in the knowledge discovery process \cite{Fay96a,Fra92}. Knowledge discovery and data mining have several applications in metabolomics which covers fields including drugs design, disease diagenesis, plant biology, environmental studies, nutrition, animal breeding, genetic studies and many other. Examples for these applications are reported in \cite{Rou10,Tay02,Wis08b,Ell06,Eno03,Fra00a,Ban19d,Wis08a}.

The applications presented in this paper aim to demonstrate MeKDDaM process model's fulfilment of the requirements of metabolomics data mining and its ability to achieve various types of analytical goals and research questions and provide coverage for metabolite profiling, target analysis, and metabolic fingerprinting. The data analysed in these applications were captured by chromatographic separation and mass spectrometry technique (LC-MS), Fourier transform infrared spectroscopy (FT-IR), and nuclear magnetic resonance spectroscopy (NMR) which involve the analysis of plant, animal, and human samples.

The applications were selected to provide coverage of both data-driven and hypothesis-driven data mining approaches and demonstrate the ability of the process to fulfil a number of data mining goals including prediction, description and verification. The selected applications are used to demonstrate MeKDDaM's ability to perform a range of data mining tasks including classification, segmentation, hypothesis testing, correlation analysis, dimensionality reduction, and feature extraction and analysis using different data mining techniques.

Each of the applications demonstrated in this research starts with a general description of the application domain. It covers the origin of the sample, the design of the assay, and goals of the original metabolomics investigation. However, some of the data are mined in a data-driven fashion, where the objectives are different from those intended in the original investigation. The results of the process applications are discussed and a conclusion is provided in Section~\ref{sec:conclsion}.

The applications were performed using MeKDDaM-SAGA \cite{Ban19c}, which is a software environment that was created in order to provide computer-aided automation, guidance and realisation for the execution of MeKDDaM process model. More details regarding the design of the software can be found in \cite{Ban19a}.

\section{MeKDDaM Process Model}
\begin{figure*}[htb!h] 
\centerline{\includegraphics[width=1\textwidth]{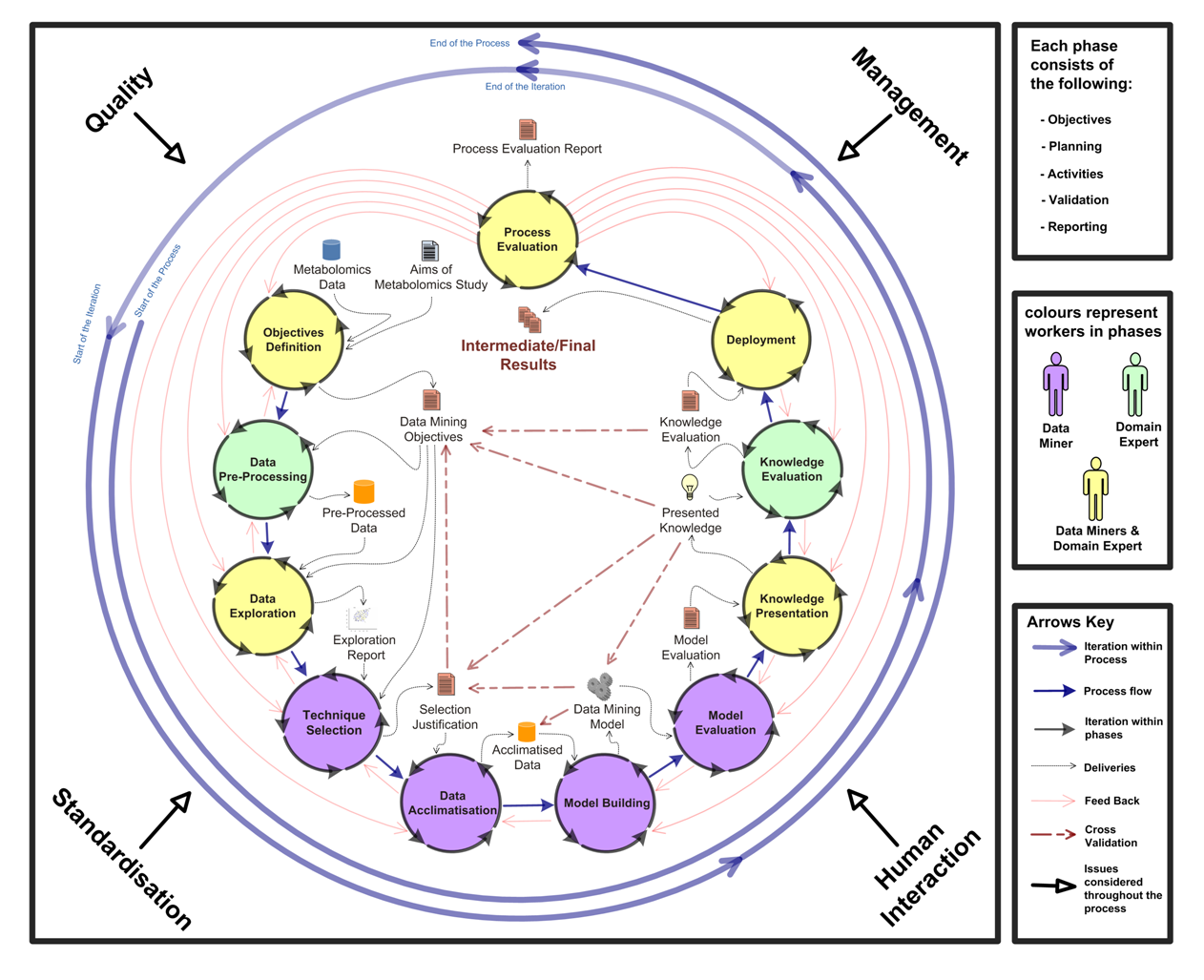}}
\centering
\caption{\textbf{MeKDDaM Process Model}-
         the graphical representation of MeKDDaM illustrates the process model structure and the flow of its phases. It also defines the inputs, deliveries and the participants in each phase.}
\label{process}
\end{figure*}

The applied MeKDDaM process model consists of eleven phases shown in Figure~\ref{process}. All process phases have the same task template, which covers their prerequisites, objectives, activities planning, performing and validation, and reporting. The process model defines its normal flow and iteration and defines feedback between its phases. The model defines the inputs to the process which cover both the metabolomics dataset and its associated meta-data in addition to the aims of metabolomics study. It also defines both the inputs and outputs of each phase.

\begin{enumerate}
\item \textbf{Objectives Definition}
provides a mechanism for defining the process objectives by matching the data mining approaches, goals, and tasks to aims of the metabolomics study and the goals of its original investigation.

\item \textbf{Data Pre-processing}
is an optional phase which aims to clean the raw data acquired by the assay instruments. The scale of this phase, and its extent and the procedures applied depends on the data acquisition instruments and techniques and on the procedures that are already performed during data acquisition.

\item \textbf{Data Exploration}
involves performing a set of activities which aim to get insight into the data through: investigation, understanding and prospecting in order to assess the potential of the data. The output of this phase takes the form of a report, which contains details regarding the activities that are performed in the phase and their outcomes.

\item \textbf{Technique Selection}
provides a strategy for selecting and justifying the selection of a data mining technique that should
achieve the process objectives and suit the nature of the targeted data. The strategy considers the requirements and feasibility of the selected technique and defines its performance measurability and success criteria. The results of this phase are the selected technique and its justification, as well as a record of factors, considerations and assessments involved. A detailed description of the strategy proposed for technique selection is available in \cite{Ban12a}.

\item \textbf{Data Acclimatisation}
involves processing and preparing the dataset(s) for \textbf{model building} and evaluation using the selected data mining technique and the applied tools. This phase generates one or more datasets, which can be used for \textbf{model building}, training and testing.

\item \textbf{Model Building}
involves building and training a data mining model that fulfills the defined process objectives by applying the selected data mining modelling technique on the dataset acclimatised in the previous phase.

\item \textbf{Model Evaluation}
involves testing, validating and evaluating the model based on the defined objectives and using measurement criteria for the technique applied. \textbf{Model evaluation} is usually performed using a separated data split which must
be allocated for model validation during \textbf{data acclimatisation} phase.

\item \textbf{Knowledge Presentation}
involves presenting the model built and validated in the previous phases in a form which presents the acquired metabolomics knowledge. \textbf{Knowledge presentation} might require performing complex visualisation techniques in order to facilitate interactive presentation of knowledge.

\item \textbf{Knowledge Evaluation}
evaluates the knowledge acquired and presented earlier from a metabolomics perspective. This
is performed in terms of its fulfillment of the objectives defined in the first phase, as well as in terms of its validity as a metabolomics knowledge, based on the background knowledge.

\item \textbf{Deployment}
aims to deploy the acquired knowledge through a mechanism that enables effective knowledge utilisation. It involves selecting appropriate deployment mechanisms in the light of the defined process objectives and within the available resources, as well as the selection of the particular deliveries which must be deployed with the knowledge.

\item \textbf{Process Evaluation} \label{subsect:Process Evaluation}
concerns evaluating the execution of the process model in terms of the flow of its phases and the validity of the tasks applied within the performed phases. It also ensure the quality of the process deliveries through the defined mechanisms of cross-deliveries validation. Cross-delivery validation is a mechanism for evaluating process deliveries. This mechanism aims at providing a high level quality assurance on the level of process deliveries. Its results can cause a feedback to an earlier phase either to resolve inconsistencies or may cause a process iteration. Possible cross-delivery validations between the process phase deliveries are shown in Figure~\ref{process}.

\end{enumerate}

All MeKDDaM process model phases allow the iterative execution of internal tasks which consist of:
\begin{itemize}
\item \textbf{Phase Prerequisites} involves confirming the validity of process inputs, phase deliveries, and other relevant information which concerns phase customisation, implementation and running. Phase prerequisites must be sufficient, specific, relevant and valid in order to perform the phase activities with no additional resources or information. This helps the justifiability and traceability of the phase results, as well as the reproducibility of its deliveries.

\item \textbf{Phase Objectives} defines the objectives which the phase is expected to achieve, the deliveries it must generate and the desired attributes and characteristics of those objectives in addition to the measurement criteria of these objectives.

\item \textbf{Phase Planning} maps the defined objectives to a set of practical actions designed to fulfil them. The planned activities must be consistent with phase prerequisites and it must comply with data mining and metabolomics procedural standards in addition to project management principles and human interaction best practices.

\item \textbf{Phase Performing} involves carrying out the phase planned activities and recording and justifying its execution. Possible problems, gaps and limitations during the phase execution must be recorded and reported as it may be used later for the purpose of phase validation, process iteration or cross-delivery validation.

\item \textbf{Phase Validation:} aims at ensuring the validity of the phase execution and the quality of its involved data and its compliance with the adopted standards. This validation is performed based on the phase planned activities and its identified objectives.

\item \textbf{Phase Reporting} concerns the generation of the phase outcomes including the tasks performed, deliveries, processed data and running report which must conform to the relevant standards in terms of their contents, format and structure.
\end{itemize}

\begin{figure}[htb!h]
\centerline{\includegraphics[width=0.9\textwidth]{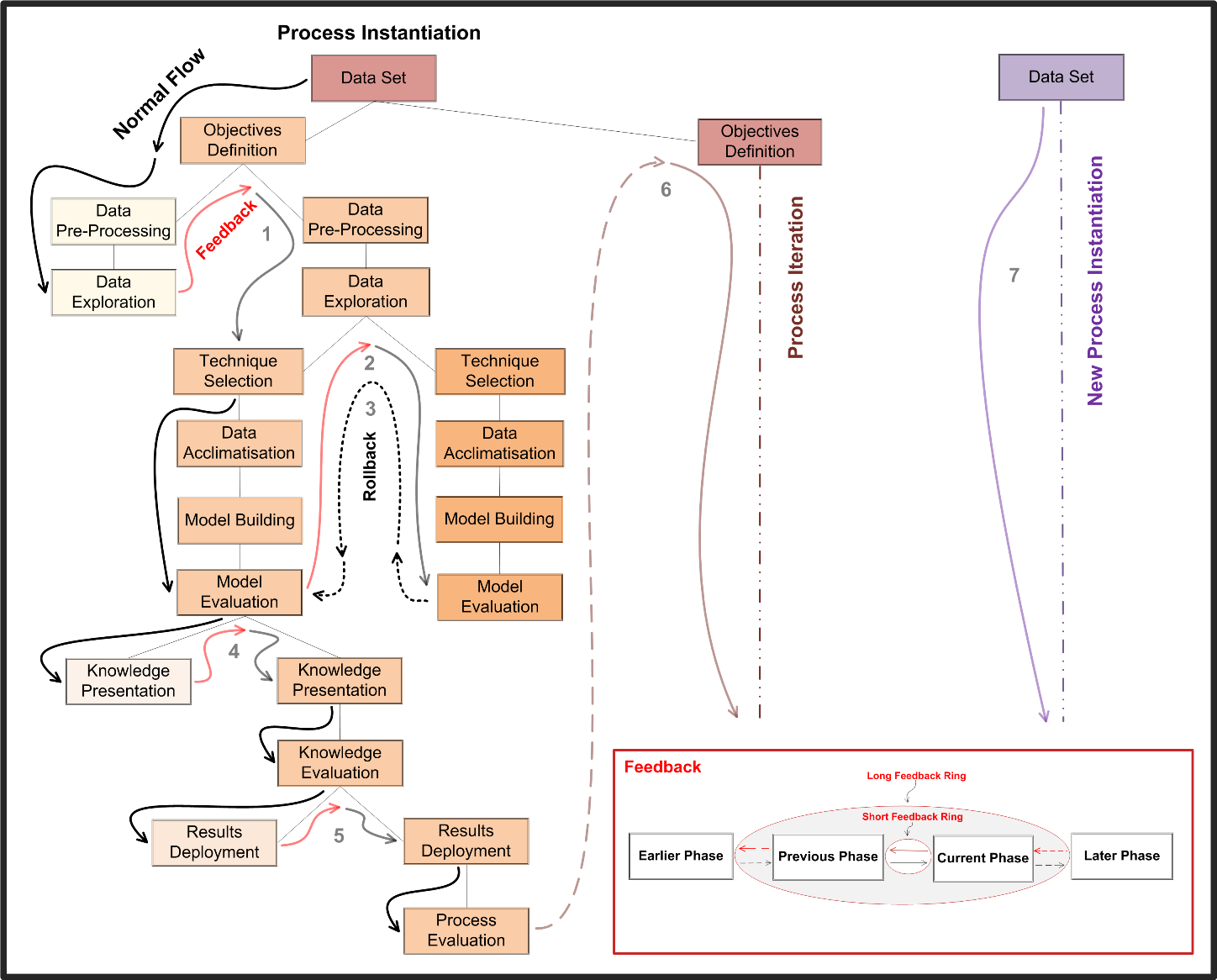}}
\centering
\caption{\textbf{Process Execution Scenarios (Repeated for Reference only:} The tree-like graph shows seven process execution scenarios, which illustrates examples of MeKDDaM process model execution that demonstrates the process feedback, rollback, phase iteration, and process iteration mechanisms.
\label{fig:ProcessExecutionScenariosCopy}}
\end{figure}

The process execution is carried out by running the internal tasks of the process phases and performing its planned activities
in order to generate its defined outcomes and deliveries either as part of the process normal flow, feedback or a process iteration.Normal flow involves executing the process phases as defined by MeKDDaM process model, while iteration involves the repetitive execution of all phases maintaining the flow of their execution as defined by the process model and into consideration the inputs and deliveries of each iteration. An iteration might be triggered due to a significant change in process objectives, or in order to formulate and test a new hypothesis, to achieve different analytical objectives, to answer a fresh or propagated question, to improve the process execution results, or to resolve major problems in the process execution. A feedback ring is a micro scale iteration that involves two or more of the process phases. It involves the re-running of all phases inside the feedback ring. Feedback is usually triggered as a response to poor phase outcomes, problems with the running of the internal tasks within the phase, or due to the inadequacy of the phase prerequisites or inputs. MeKDDaM also supports rollback, which enables undoing or discarding phase or process iteration when it fails to achieve its purpose or intended objectives e.g. building a better model. Figure~\ref{fig:ProcessExecutionScenariosCopy} provides examples for the process execution scenarios that illustrate the process normal flow as well as concepts of iteration, feedback, and rollback.

\section{\emph{Arabidopsis} Isoprenoids Application}
\label{sec:isoprenoids}

\subsection{Application Description}
This application involves applying metabolomics in plant genetics. It aimed at testing the discrimination between \emph{Arabidopsis thaliana} genotypes using samples of the plant leaf extracts in order to analyse the impact of 6 gene mutations on a group of metabolites known as isoprenoids. The mutations covered both fatty acid and lipid metabolism.

\subsection{Materials and Methods}
\subsubsection{Experimental Protocol}
The \emph{Arabidopsis} plants were grown and harvested by the HiMet project collaborators in The John Innes Centre(JIC), where the growth, harvest, extraction and preparation protocols were collected and recorded as meta-data in compliance with ArMet model \cite{Jen04}. The plants were grown in a constant environment using a block design to control for environmental variation. The harvesting protocols involved collecting aerial tissues from the plants. The metabolism was then quenched using liquid nitrogen and then frozen, dried, and powdered. They were then shipped at ambient temperature within a few days and otherwise stored at a temperature of -80\textcelsius. The samples were analysed using LC-MS instrument based on protocols described in \cite{Fra00a} with modifications. $\alpha$-tocopherol was used as an internal standard. The quantification of the isoprenoids was conducted using positive-ion atmospheric pressure chemical ionisation MS \cite{Sco10}.

\subsubsection{Data Acquisition}
 The dataset used in this application was originally acquired by York University as part of the HiMet Project. The instances in the acquired dataset represent samples of 14 mutants in addition to a wild type. Each of the plant genotypes was grown in 9 blocks or shelves and were analysed using 3 replicates. The total number of samples was 399, as 6 samples were missing (3 sample replicas x 2 mutant blocks).

\subsection{Results of MeKDDaM Process Model Execution}
The process execution considered recorded the management aspects of the process execution including cost and time constraints as well as the process resources including software, hardware and human expertise interaction. The application encouraged the utilisation of both data mining and metabolomics reporting standards, while the quality assurance policy stated that the percentage of missing data should not exceed 10\% of the targeted data and the classification error rate should not exceed 40\%. Process execution proceeded through customized phases as follows:

The \textbf{objectives }of the process have been defined in a hypothesis-driven fashion, where the study was designed based on the hypothesis that \emph{Arabidopsis} mutations related to a particular area of metabolism, can be monitored through analysing the plant isoprenoids profile. Therefore, the process objectives were set to classify the samples into three areas of metabolism related to the \emph{Arabidopsis} genotypes described earlier in the section. The traceability mechanism provided by the software environment \cite{Ban19a,Ban19c} was used to document the bases on which the objectives were derived.

The metrics for assessing the fulfilment of the objectives were also defined as well as their success criteria. The threshold for the acceptable classification accuracy was set to 60\% and achieving a satisfactory ratio of true positive predictions across the three predicted classes. The objectives were also assessed in terms of their feasibility based on the process constraints and the project available resources. The potential fulfilment of the objectives was also assessed according to the dataset sufficiency, adequacy, and relevance to fulfil the process objectives.

Most of the required \textbf{data pre-processing} procedures had been performed already during data acquisition, One particular procedure was identified as still required. Some of the values recorded in the profile as zeros, where in fact under the detection limit of the machine, the thing which might introduce bias to the data as well as influence the successive \textbf{data pre-processing}, exploration, and acclimatisation phases or even affect the performance of the data mining model as these values might be treated as actual values in the calculation rather than their original meaning in the context of the machine detection limit. As this problem is common in metabolomics data analysis, several strategies are available for tackling it, such as replacement by mean, median, or minimum. In this application, we have chosen to replace these zero values by the mean of the machine replicates within the same biological sample. This solution was expected to resolve these issues as the machine replicates were originally designed to resolve the instrumental bias introduced during the data acquisition.

\begin{figure}[htb!h] 
\centerline{\includegraphics[width=0.9\textwidth]{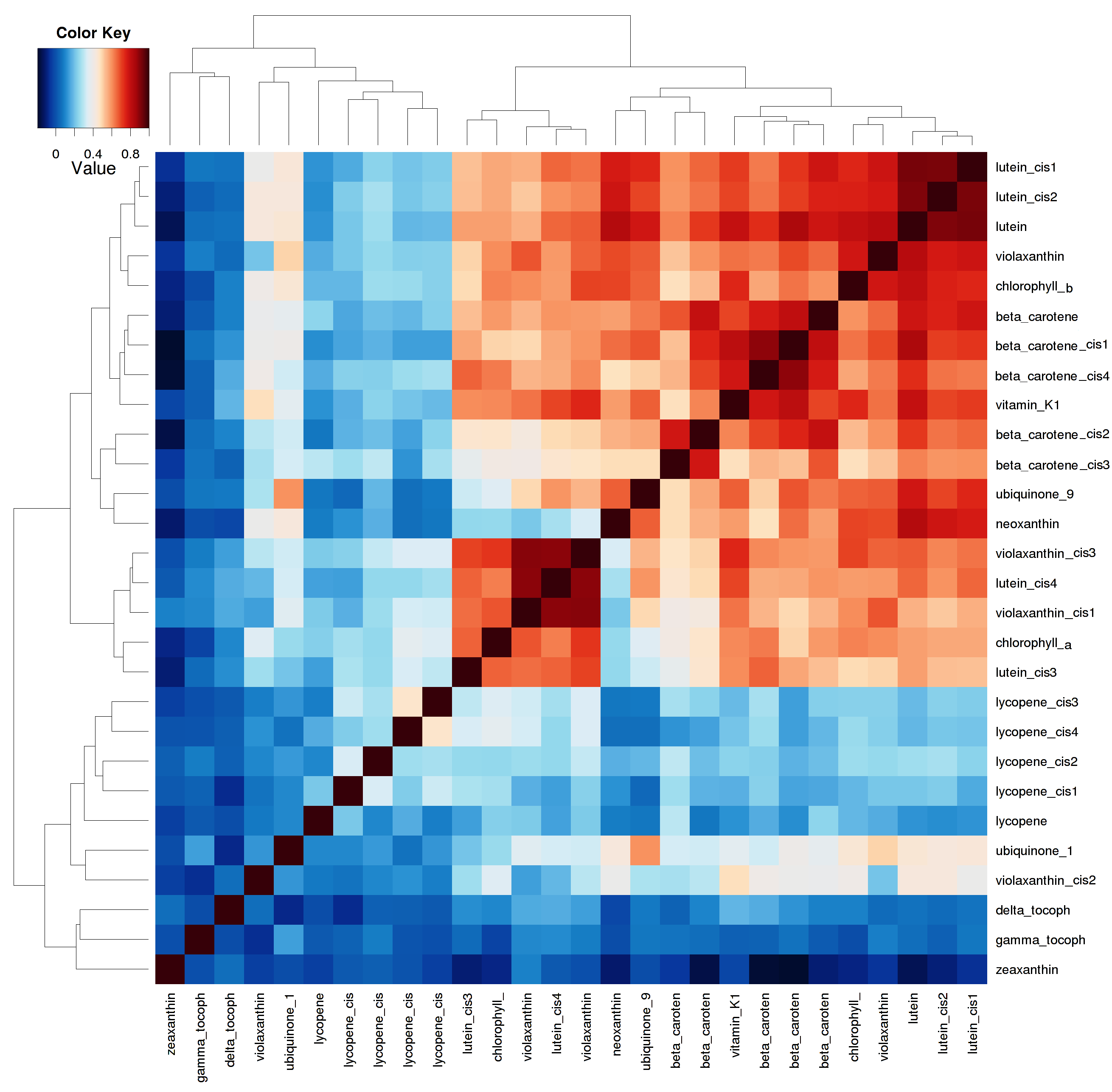}}
\centering
\caption{\textbf{A heat map illustrates the correlation between the metabolites in the profile:} Red colours represents positive correlations, while blue represent negative correlation. The darkness of colour indicates the strength of the correlation. \label{fig:HiMet_data_exploration_prospecting}}
\end{figure}

\begin{figure}[htb!h] 
\centerline{\includegraphics[width=0.9\textwidth]{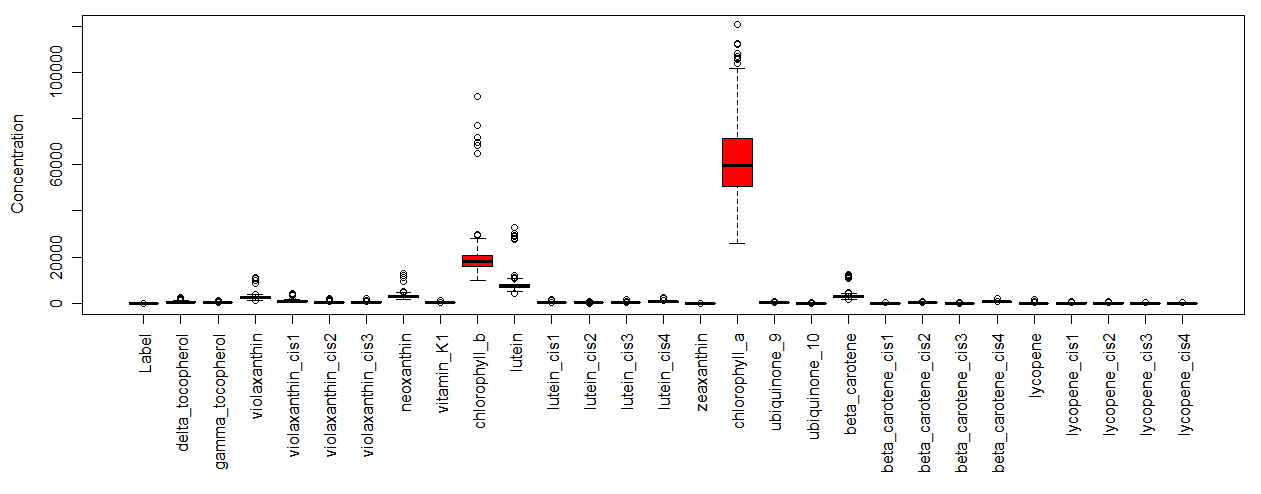}}
\centering
\caption{\textbf{Distribution of the Data in the \emph{Arabidopsis} Isoprenoids Application:}A box plot showing the distribution of the data and basic statistics about the values in its features \label{fig:HiMet_data_exploration_raw_BoxPlot}}
\end{figure}

During \textbf{data exploration} phase, it was decided that a box plot would be appropriate for prospecting the data trends and values. However, the data understanding uncovered the existence of an attribute which corresponds to an internal standard, which was described in the study. Therefore, feedback from \textbf{data exploration} to \textbf{data pre-processing} phase was decided in order to iterate through the \textbf{data preprocessing} phase and perform a column-wise scaling in reference to the internal standard, which was seen as another mean of reducing the machine variation. This is indeed an illustration of the advantageous of the feedback mechanism provided by MeKDDaM process model as it enabled improving the quality of the preprocessed data. Furthermore, the \textbf{data exploration} phase was re-run again in order to reflect the changes in the dataset using the same procedures, which have been performed earlier. The correlation between the profile metabolites is prospected in Figure~\ref{fig:HiMet_data_exploration_prospecting} using a heat map, while the investigation of the data distribution and outliers is illustrated by the box plot in Figure~\ref{fig:HiMet_data_exploration_raw_BoxPlot}. The heat map illustrates a relatively strong correlation between violaxanthin, chlorophyll, and lutein as well as between vitamin K1 and $\beta$-carotene, and between neoxanthin and lutein.

The\textbf{technique selection} phase was then performed based on the phase predefined tasks as described by the process model and as customised by the current execution. Artificial neural networks (ANN) was seen as a possible classification technique which has the potential of the achieving the process objectives as well as suiting the nature of the data as ANN is capable of classifying data with numeric attributes and categorical classes, which matches the process objectives and the nature of the data in metabolite profile. The technique specific measurability was also defined as well as its success criteria. The \textbf{technique selection} was also assessed in terms of technical availability and feasibility. Justifications of its selection were also recorded.

\begin{figure}[htb!h]
\centerline{\includegraphics[width=0.9\textwidth]{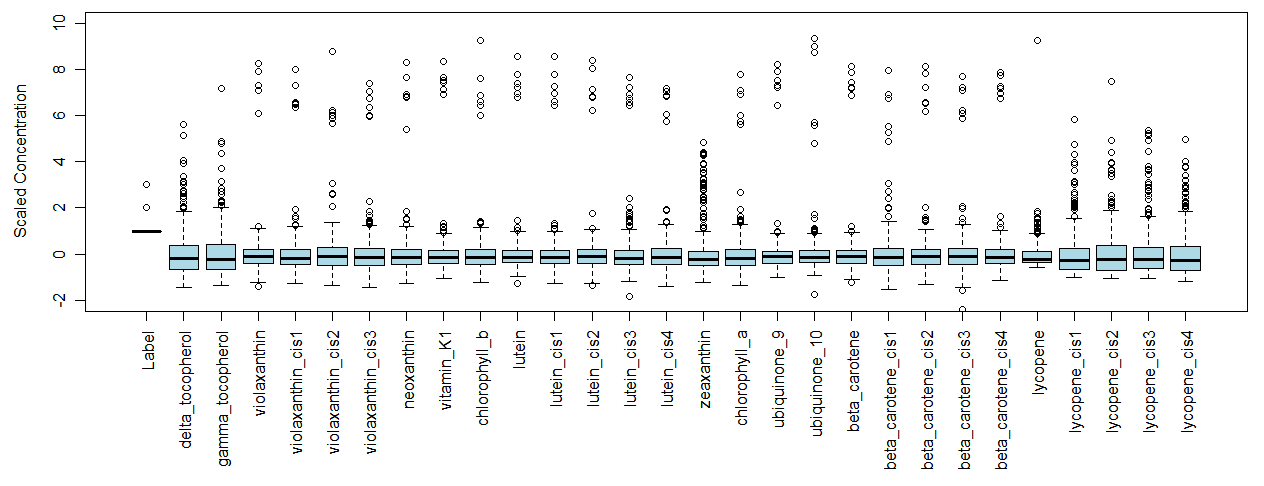}}
\centering
\caption{\textbf{Distribution of the Normalised Data in the \emph{Arabidopsis} Isoprenoids Application:} A box plot showing the normalised data after it was scaled with auto-scaling. \label{fig:HiMet_data_exploration_normaised_BoxPlot}}
\end{figure}

The \textbf{data acclimatisation} phase was then executed based on the outcomes of \textbf{data exploration}. The selected technique requires normally distributed data and it was found skewed to the left. The selection of auto-scaling method was based on \cite{Van06}, who reported the successful application of the auto-scaling in normalising similar chromatography data. Auto-scaling was successful in normalising the data distribution as illustrated in the box plot shown in Figure~\ref{fig:HiMet_data_exploration_normaised_BoxPlot}. The data were then randomised to obtain a randomised distribution of the samples in the different data splits and the data were then divided into two splits. The first was allocated for model training with a proportion of 60\% of the original data, while the remaining data was allocated for model testing. This proportion was selected in the light of the availability of sufficient data for performing both model training and testing and based on the common practices in machine learning techniques validation and testing as reported in data mining literature \cite{Han01,Kan03,Mai05a,Wit05}.

In the \textbf{model building phase}, the model was trained based on 500 epochs and the performance of the model was automatically generated. The model, together with a report on the parameters of its generation and its performance was the output of this phase.

\begin{figure}[htb!h]
\centerline{\includegraphics[width=0.5\textwidth]{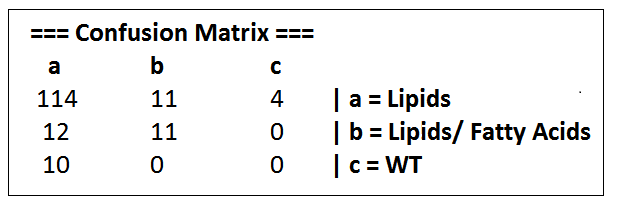}}
\centering
\caption{\textbf{Evaluation of the MLP Model in the \emph{Arabidopsis} Isoprenoids Application:} A confusion matrix showing the prediction accuracy of the MLP model
\label{fig:HiMet_MLP_model_evaluation}}
\end{figure}

The \textbf{model evaluation} phase was performed based on the measurable success criteria that were set in the objective definition phase. The error rate was 23.0\%, while the percentage of the correctly classified samples was 77.0\%. The confusion matrix of the \textbf{model evaluation} is shown in Figure~\ref{fig:HiMet_MLP_model_evaluation}, while the details of the \textbf{model evaluation} are shown in a persisted XML file.

Feedback from \textbf{model evaluation} phase to \textbf{technique selection} was decided because of the lack of correctly classified samples in the wild type class (WT) as the success criteria required achieving a satisfactory ratio of true positive prediction across the three predicted classes. The \textbf{technique selection} phase was then iterated in order to select a different modelling technique. Decision trees were selected as an alternative data mining technique was performed based on the \textbf{technique selection} strategy described by the process model, where it was matched as a potential technique for achieving the process objectives while suiting the nature of the data in terms of its attributes data type and the ratio between the number of data samples and attributes. A detailed description of the strategy proposed for technique selection is available in \cite{Ban12a}.

The \textbf{data acclimatisation} phase was then executed, where the data was acclimatised using the same procedures applied earlier (auto-scaling, randomisation, and data splitting), which were applied in order to make the data suitable for \textbf{model building} and evaluation using C4.5 as an alternative data mining technique.

\begin{figure}[htb!h]
\centerline{\includegraphics[width=0.5\textwidth]{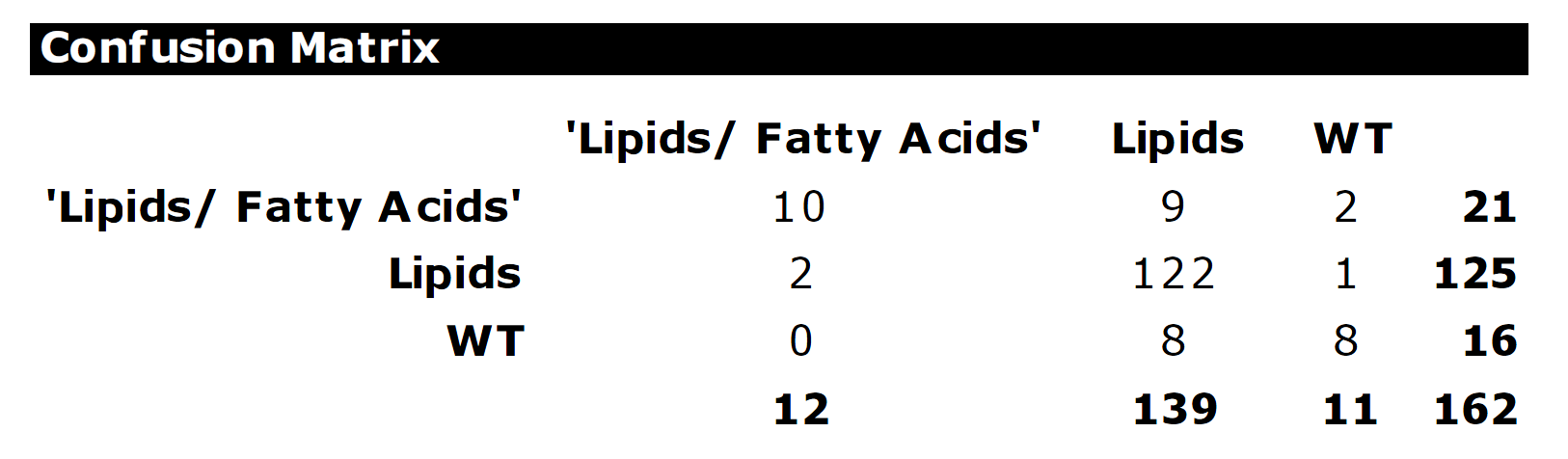}}
\centering
\caption{\textbf{Evaluation of the C4.5 in the \emph{Arabidopsis} Isoprenoids Application:} A confusion matrix Showing the prediction accuracy of the C4.5 model.
\label{fig:HiMet_C45_model_evaluation}}
\end{figure}

The \textbf{model building} phase was then executed. The C4.5 decision tree model was built and trained using the data allocated for \textbf{model building} and training. C4.5 model was able to classify 86.42\% of the samples correctly, which was better than the rate achieved by the MLP model. The distribution of the correctly classified samples in the C4.5 confusion matrix was also better than that of the MLP model. Therefore, the C4.5 model satisfies the success criteria defined by the process objectives, with a prediction accuracy over 60\% and with a satisfactory true positive prediction ratio across all of the prediction classes. The confusion matrix of the C4.5 \textbf{model evaluation} is illustrated in Figure~\ref{fig:HiMet_C45_model_evaluation}.

As the results of the \textbf{model evaluation} were satisfactory, the model was forwarded to \textbf{knowledge presentation} phase, where the model was presented using the decision trees graphical representation as shown in Figure~\ref{fig:HiMet_C45_model_visualisation}. Each of the prediction classes is represented in the tree in a different colour. The darkness of the colour represents the confidence of the class prediction, which is also represented by the real number and the small pie chart on each of the tree nodes, while the integer number on each node represents the total number of the samples, which support the class prediction.

\input{himet_c45_model_visualisation.tex}

The results were then evaluated by running the \textbf{knowledge evaluation} phase. The results fulfilled the defined process objectives which aimed to classify the samples into the three \emph{Arabidopsis} areas of metabolism. The classification results were found valid according to the domain expert background knowledge and experience and in terms of its consistency with the biological and biochemical principles related to metabolomics.

The results have been found consistent with the results reported by similar studies \cite{Fra00a,Sco10}. Zeaxanthin, vitamin k1, chlorophyll, violaxanthin, lycopene, and $\delta$-tocopherol were found important for classifying the samples as illustrated by the decision tree diagram in Figure~\ref{fig:HiMet_C45_model_visualisation}. However, some of these metabolites such as violaxanthin, vitamin k1, and chlorophyll, were found having a strong correlation with other metabolites as discussed earlier in \textbf{data exploration} and as illustrated by the heat map in Figure~\ref{fig:HiMet_data_exploration_prospecting}.

The \textbf{deployment} phase was executed using the available file system \textbf{deployment} mechanism for the purpose of results publication. The process execution was evaluated in terms of its consistency with the process model layout and in terms of its consideration of the process practical aspects. The human interaction was consistent with the roles which have been defined by the process model. The quality of the process involved data, procedures and reports was also satisfactory and meets the defined process policy. The quality assurance policy stated that the data must have no more than 10\% of missing data, which must be handled during \textbf{data acclimatisation} and should have no outliers which could affect the results of \textbf{model building} and evaluation. It also stated that the model performance must be satisfactory across all classes and that all the process procedures must be recorded and justified based on traceable evidence. The policy also stated that all the process deliveries must be reported during the process \textbf{deployment}.

The reporting of the process execution and deliveries was performed according to the defined standard policy, which was based on both Metabolomics Standards Initiative (MSI) \cite{Har07b,Fie07} which defines the minimum reporting for metabolomics studies and data mining reporting standards e.g. Predictive Model Markup Language (PMML) \cite{Kur06,Pec06} which defines an interchangeable standard format for representing data mining models.

The cross-delivery validation was also performed on the data mining objectives report generated in the first phase, selection justification report generated in the fourth phase, the data acclimatised in the fifth phase and the data mining model built in phase six. The validation confirms these reports correctness, completeness and consistency with the deliveries of the later phases as defined in MeKDDaM process model.

\section{Cow Diet Application}\label{sec:CowDiet}
This application involves applying metabolomics in dairy management systems which aims to enhance milk productivity and quality. The dataset in this application was originally generated by University of Alberta, Edmonton, Canada \cite{Ame10}.

\subsection{Application Description}
The investigation aimed to analyse the changes in dairy cow rumen metabolites under a controlled diet using NMR profiling. The study was designed to track the changes in rumen metabolites over a period of time using different cow serial diet.

\subsection{Materials and Methods}

\subsubsection{Experimental Protocol}
 The study was designed to track the changes in rumen metabolites over a period of 21 days. The first 11 days were used to calibrate the cows' diet, while the last 10 days were used for sample collection across five-time intervals in days 1,3,5,7 and 10 of the last 11 days of the study. The samples were collected from rumen fluid of eight healthy cows, where two cows were randomly allocated to each of the four diets. Each group of the cows, were fed a different percentage of barley that represent 0\%, 15\%, 30\%, and 45\% of their diet dry matter. The samples were replicated based on 4 a \emph{X} 4 Latin square design. More details regarding the study design are available in \cite{Ame10}.

\subsubsection{Data Acquisition}
The dataset was acquired using $^{1}_{}H$-NMR on 500MHz frequency. NMR spectra were preprocessed by a dedicated package which is called Chenomx NMR Suite 6.0. In each 64 k of the frequency domain, the FIDs were zero filled and multiplied by the exponential weight of the line broadening the value of 0.5 Hz. Fourier transformation was then applied on NMR spectra which was referenced to 0.0 ppm. The spectra baseline was then corrected and phased manually. The metabolites have been profiled by quantifying the concentration of the compounds identified in the spectra \cite{Ame10}.

\subsection{Results of MeKDDaM Process Model Execution}
The process inputs consist of the acquired dataset and the aims of the study as defined by MeKDDaM process model in Figure~\ref{process}, while the \textbf{objectives} of the analysis were set to classify the samples into the four dietary groups using the NMR metabolite profile. In addition, the defined objectives measurability, feasibility, and success criteria were also defined.

No \textbf{preprocessing} was required, as the data was already processed during its data acquisition stage. This was carried out using the software system integrated into the data acquisition instruments. The profile was normalised by dividing the concentration of the measured metabolites into the total concentration of all metabolites in that sample, while the metabolite identifications were confirmed by sample spiking which was used to confirm compound spectral signatures by adding presumptive compounds to the samples \cite{Ame10}. However, the preprocessing phase was formally reviewed in order to confirm the validity of its performance and the completeness, correctness, and consistency of its delivery.

\begin{figure}[htb!h]
\centerline{\includegraphics[width=0.9\textwidth]{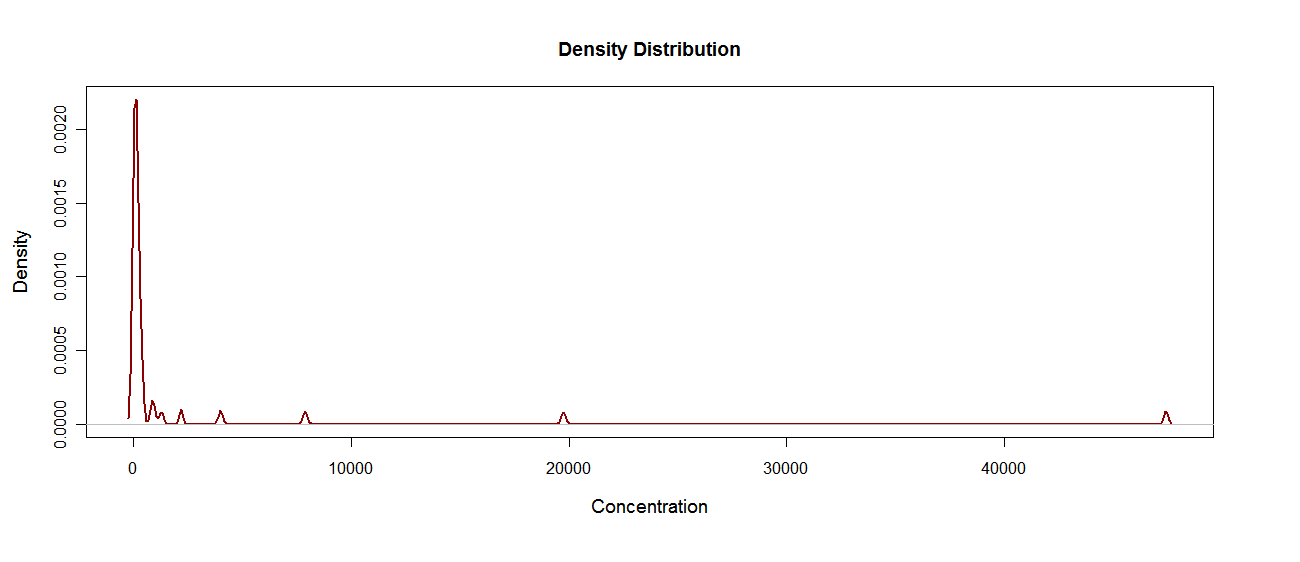}}
\centering
\caption{\textbf{Distribution of the Data in the Cow Diet Application:} A density plot generated by R\cite{Cra07,R10} that represents a density function which combined the distribution of all samples in the profile.
\label{fig:cow_raw_data_distribution}}
\end{figure}

\textbf{Data exploration} showed that the data included 49 attributes and 39 samples. The first attribute has nominal values which represent the samples id, while the second attribute holds values which correspond to the dietary classes. The remaining 47 attributes hold continuous numerical values which represent the metabolite concentrations. The distribution of the metabolite concentrations was investigated as well as the quality of the data covering the existence of outliers and missing values. A density function was plotted to represent the samples overall distribution, which revealed skewness in the data distribution towards the left as illustrated in Figure~\ref{fig:cow_raw_data_distribution}, while no missing values or significant outliers were detected. However, the data included 2.8\% of zero values.

\begin{figure}[htb!h]
\centerline{\includegraphics[width=0.9\textwidth]{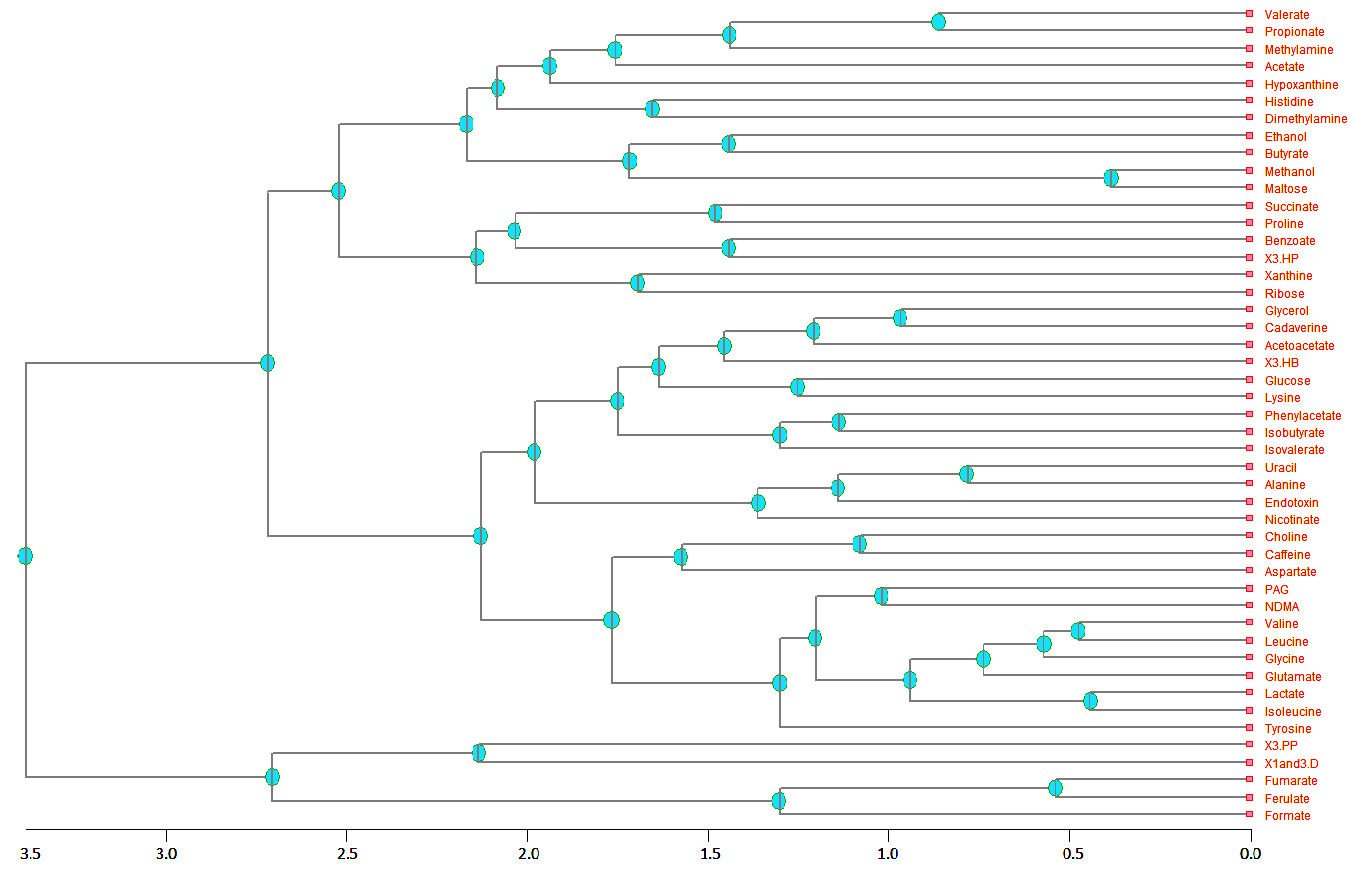}}
\centering
\caption{\textbf{Correlation of the Data in the Cow Diet Application:} A dendrogram showing the correlation between the targeted metabolites in the NMR profile. The correlation was found using Pearson hierarchical correlation
\label{fig:cow_raw_data_correlation}}
\end{figure}

The correlation between the data attributes was also prospected, which uncovered significant correlation between some of the profile metabolites such as between valine, leucine, glycine, and glutamate as well as between some pairs of metabolites including methanol and maltose, lactate and isoleucine, fumarate and ferulate, valerate and propionate, and between alanine and uracil. Figure~\ref{fig:cow_raw_data_correlation} illustrates the correlation between targeted metabolites using Pearson hierarchical correlation. These typical outputs from \textbf{data exploration} were recorded as output from the phase and hence input to \textbf{technique selection}.

The tasks of the \textbf{technique selection} phase were performed according to the customised process description. The selection of the technique considered the defined objectives as well as the nature of data. The process objectives were matched to data mining discovery approach and its prediction goals, which can be achieved through a number of data mining tasks which include: segmentation, regression and classification. Segmentation was excluded as the data needs to be classified to the pre-known set of classes rather than naturally occurring groups. On the other hand, regression techniques were excluded as due to the prediction classes data types. This narrowed the \textbf{technique selection} to the supervised classification techniques.

However, few of these techniques were found to suit the nature of the data, as the attributes outnumber the samples, which makes decision trees unsuitable. On the other hand, the small number of samples makes the performance of a neural network model unreliable.

PLS-DA and random forest were both seen as possible techniques, which might suit the nature of the data and have the potential to achieve the process objectives. Additionally, the two techniques are feasible to perform within the application time and cost constraints In addition to the availability of hardware, software resources and human expertise. However, no particular preference made one of these techniques a better choice than the other, and therefore the choice between them was left to be decided based on their classification performance. So, PLS-DA was reported as initial selection.

\begin{figure}[htb!h]
\centerline{\includegraphics[width=0.9\textwidth]{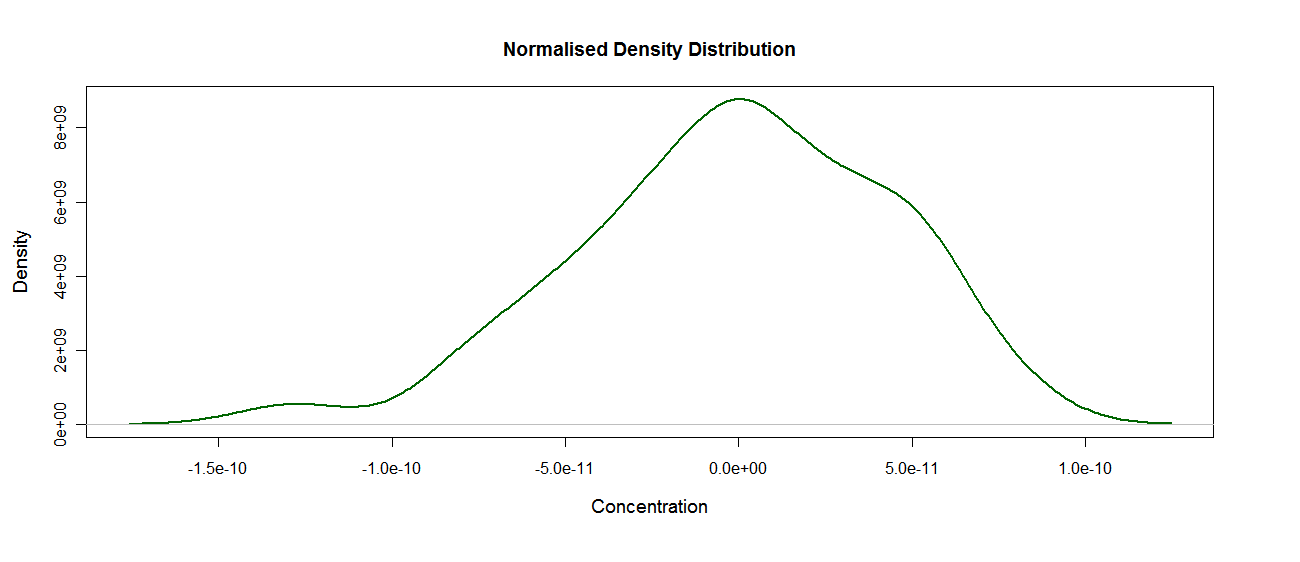}}
\centering
\caption{\textbf{Distribution of the normalised data in the Cow Diet Application:} A density plot generated by R \cite{Cra07,R10} that represents a density function which combined the distribution of the samples in the normalised data
\label{fig:cow_normalized_data_distribution}}
\end{figure}

\textbf{Data acclimatisation} was performed based on the requirements of the selected technique (PLS-DA) as well as in the light of the defined process objectives and the \textbf{data exploration} report. The data distribution was normalised using auto-scaling. The density plot in Figure~\ref{fig:cow_normalized_data_distribution} shows the data distribution after normalisation. The dataset was then randomised and split into two equal proportion as the number of samples is quite limited. The two datasets and the originally acclimatised datasets were all stored in formats that are accessible by the applied \textbf{model building} software.

\begin{figure}[htb!h]
\centerline{\includegraphics[width=0.5\textwidth]{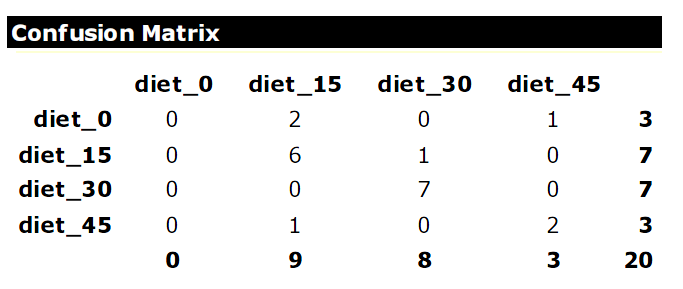}}
\centering
\caption{\textbf{Evaluation of the random forest Model in the Cow Diet Application:} A confusion matrix showing the prediction accuracy of the random forest model.
\label{fig:cow_raw_data_RF_model_evaluation}}
\end{figure}

The \textbf{model building} phase was carried out using the training data, while the testing split was used for validating the model. The model was able to classify samples, but with a limited prediction accuracy, which was just above the 60\% success criteria set in process objectives. Therefore, the decision was made to feedback to the model selection phase and select random forest as an alternative technique. The random forest model required no additional acclimatisation other than those, which have already been performed for PLS-DA \textbf{model building}. The random forest model was built using Orange and performed relatively better than the one which was built using PLS-DA as it achieved a prediction accuracy rate of 75\%. The confusion matrix of the \textbf{model evaluation} is shown in Figure~\ref{fig:cow_raw_data_RF_model_evaluation}.

\begin{figure}[htb!h]
\centerline{\includegraphics[width=0.9\textwidth]{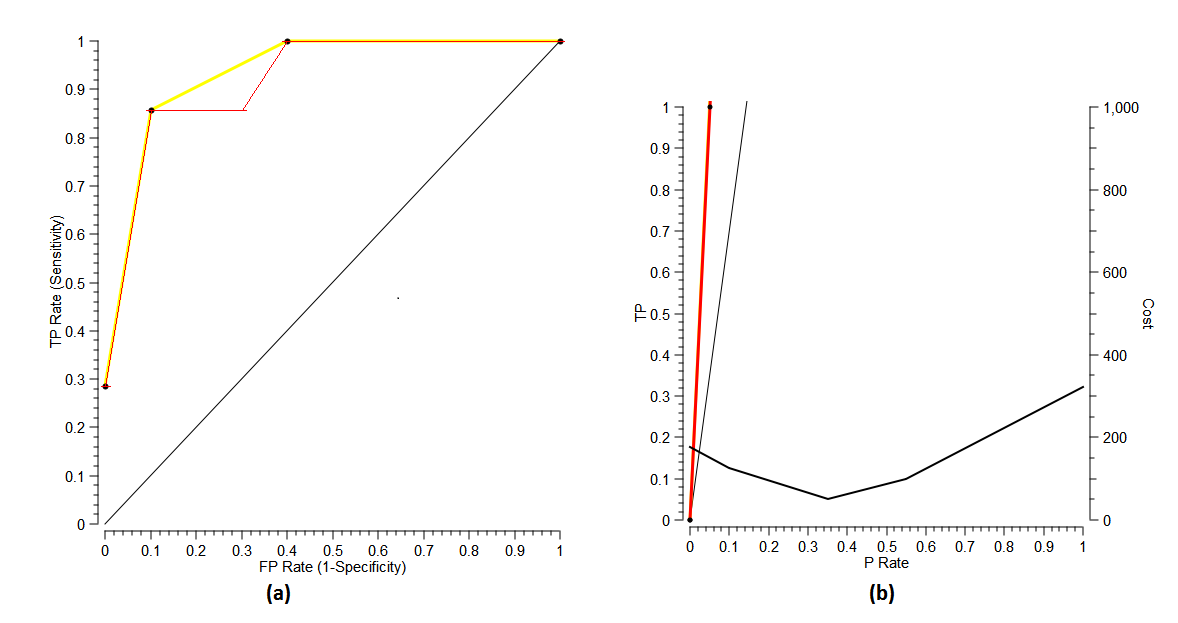}}
\centering
\caption{\textbf{The random forest \textbf{model evaluation} using ROC and Lift charts:} (a) ROC chart. The red curve represents the relationship between the true positive rate and false positive rate, and the yellow curve represents the ROC convex hull and provides an indication regarding the model sensitivity. (b) Lift chart. The red curve represents the relationship between true positive and positive values, while the black curve represents the cost function.
\label{fig:cow_raw_data_RF_model_evaluation_ROC_Lift}}
\end{figure}

Due to the poor visualisation of random forest in general, the knowledge was presented in a textual format, which describes both the model and its evaluation results. The \textbf{model evaluation} was visualised using ROC and lift charts, which are shown in Figure~\ref{fig:cow_raw_data_RF_model_evaluation_ROC_Lift}. These charts are useful for presenting the \textbf{model evaluation} and assessing its sensitivity and precision. More details regarding these methods of evaluation were discussed earlier. ROC provides insight into the relationship between the true positive and false positive rates, while Lift chart is useful for analysing the relationship between the true positive rate and the total of both true and false positive rates. Due to the better performance achieved by the random forest model, the new model was selected to be forwarded to the successive phases in the process.

The knowledge was presented by the random forest model and its performance evaluation was evaluated as a successful fulfilment of the defined process objectives and its performance exceeds the defined success criteria which were set to 60\% classification accuracy. However, the results propagated a question regarding the identification of the profile metabolites which are responsible more for the difference in the diet classes. This might lead to additional useful knowledge regarding the correlation between the change in dairy cows diet and some of these metabolites.

The results of the model, which included all the process deliveries, were deployed using the file system in order to be reported in this publication. The \textbf{process evaluation} validated the process execution in terms of its consistency with the process layout structure and the project resources. It also validated the process consideration of the process of practical aspects and performed cross-delivery validation based on the mechanism defined by MeKDDaM process model and implemented by the process realisation and automation software \cite{Ban19a,Ban19c}. The cross-delivery validation results confirmed the validity of the process \textbf{objectives definition}, \textbf{technique selection} justification, reported \textbf{data acclimatisation} procedures and hence the validity of the model.

The process was iterated to answer the new propagated question where the objectives were set to finding the metabolites which are significant in discriminating the samples in the four diet class. The \textbf{data pre-processing} and \textbf{data exploration} phases were revisited. The new process objective did not instigate any changes in \textbf{data pre-processing}. However, \textbf{data exploration} was enhanced to give more insight regarding the relationship between metabolites which had shown a strong correlation. The correlation was re-prospected showing the correlation between each pair of profile metabolites using a scatter plot.

\begin{figure}[htb!h]
\centerline{\includegraphics[width=0.5\textwidth]{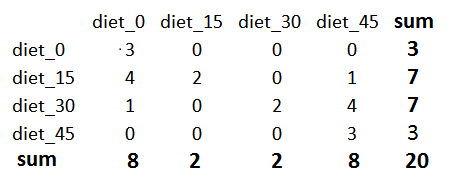}}
\centering
\caption{\textbf{Evaluation of the Random Forest Model in the Cow Diet Application:} A confusion matrix showing the prediction accuracy of the random forest model in order to find the important variables in the second iteration.
\label{fig:cow_raw_data_RF_vip_model_evaluation}}
\end{figure}

The \textbf{technique selection} was re-executed, where the random forest was selected as the technique for performing feature extraction and analysis since it matches more the new process objectives which focus on variables scoring as much as on samples classification. Consequently, \textbf{Data acclimatisation} phase was also revisited, but no additional acclimatisation procedures were needed. The \textbf{model was then built} and then evaluated. The evaluation results showed that the model did not only provided extra information regarding the variable importance score, but it also archived better prediction accuracy across all classes. The confusion matrix of the random forest \textbf{model evaluation} is illustrated in Figure~\ref{fig:cow_raw_data_RF_vip_model_evaluation}.

The \textbf{knowledge presentation} phase was re-run through its phase iteration mechanism since Rattle provides visualisation for the variable importance. The \textbf{knowledge evaluation} phase was also re-run in order to evaluate the new knowledge in the context of metabolomics, where it was performed in a similar fashion to the first iteration.

\begin{figure}[htb!h]
\centerline{\includegraphics[width=0.9\textwidth]{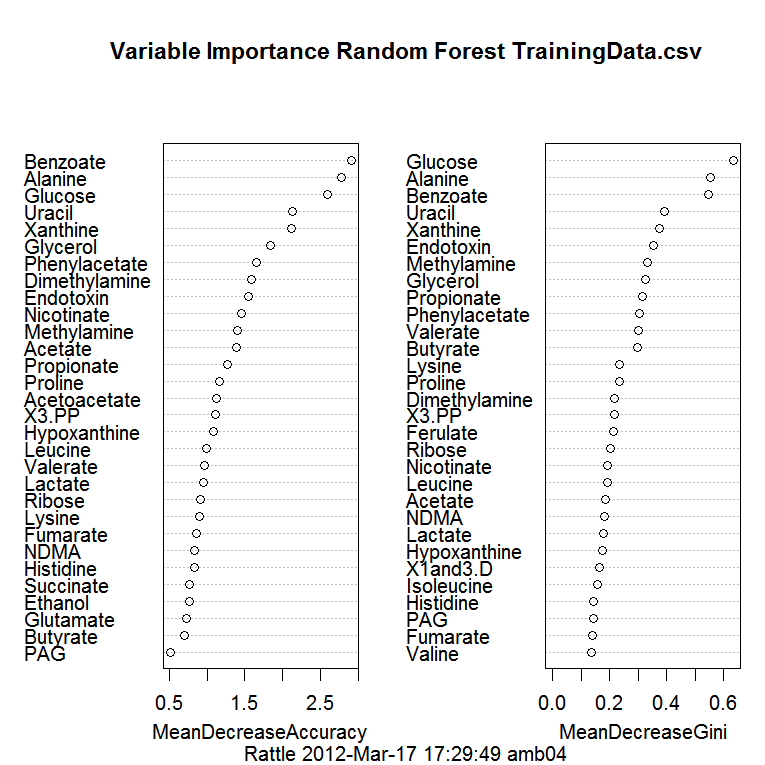}}
\centering
\caption{\textbf{Visualisation of the Variable Importance in the Cow Diet Application:} An illustration of the important features found in the cow diet data using random forest model. The variable importance was computed using both mean decrease accuracy, and mean decrease Gini index methods, which are described in \cite{Wil11}.
\label{fig:cow_raw_data_RF_vip_model_visulaisation}}
\end{figure}

The results were validated by the domain expert based on the known biological and biochemical principles. The results in this iteration were comparable to the results produced by \cite{Ame10}, who used the same set to answer a similar question which aimed to find the metabolites which their concentration was influenced more by the change in the proportion of the barley diet. However, the results of the reported analysis were generated using ANOVA and PCA, which are both sensitive to the techniques which are used for \textbf{data acclimatisation} such as scaling. Figure~\ref{fig:cow_raw_data_RF_vip_model_visulaisation} shows the important variables, which contributed more to the difference between the four diet classes.

The important metabolites highlighted by the random forest model are similar to those reported by \cite{Ame10}, who reported an increased level of the concentration of methylamine, glucose, alanine, maltose, propionate, uracil, valerate, xanthine, ethanol, and phenylacetate. However, while the results of the random forest model rank all of these metabolites as important variables apart from ethanol, it also ranks benzoate, glycerol, endotoxin, dimethylamine and nicotinate as important variables as shown in Figure~\ref{fig:cow_raw_data_RF_vip_model_visulaisation}.

As the results of the process second iteration were found valid, the process was terminated and the knowledge and its associated deliveries were deployed as final results as the model fulfilled the objectives set in the first and second iteration of the process regarding samples classification as well as variable importance. Finally, the process execution was evaluated in a similar fashion to the first iteration and found to be valid as well as the cross-delivery validation which ensured the validity of the deliveries of the process iteration.

\section{\emph{Arabidopsis} Fingerprinting Application}
Similar to the previous application, the metabolomics investigation in this application involves the study of plant genetics and their area of metabolism. It aims to analyse the impact of \emph{Arabidopsis thaliana} gene knock-out using metabolic fingerprint data. The dataset in this application was acquired by HiMet project collaborators in Aberystwyth University \cite{Sco10}.

\subsection{Application Domain Description}
The materials in this application were the same as the materials in Section~\ref{sec:isoprenoids}. However, the sample preparation procedures are quite different, as FT-IR was used for acquiring metabolic fingerprinting data \cite{Sco10}.

\subsection{Materials and Methods}
\subsubsection{Experimental Protocol}
The materials in this application were the same as the materials in Section~\ref{sec:isoprenoids}. However, the sample preparation procedures are quite different, as FT-IR was used for acquiring metabolic fingerprinting data \cite{Sco10}. The sample materials were analysed by HiMet project collaborators at Aberystwyth University using FT-IR spectrometer.

\subsubsection{Data Acquisition}
The IR absorbance spectra were captured and recorded between 4,000$cm^{-1}$ and 600 $cm^{-1}$ at resolution of 4 $cm^{-1}$, with 256 spectra averaged per sample. The averaged duplicate plate spectra consisted of 1,764 variables with nonzero values. The acquired data was preprocessed by normalising to zero mean and to a unit standard deviation.

\subsection{Results of MeKDDaM Process Model Execution}
The process execution starts with importing the FT-IR metabolic fingerprint data and by recording the application information regarding the design of the study, which was used for acquiring the data in a similar fashion to that described in the \emph{Arabidopsis} isoprenoids application.

The \textbf{objectives} of the process was set in a data-driven fashion, where it was set to classify the samples into the five areas of metabolism which are linked to a group of \emph{Arabidopsis} genotypes. The objectives were justified using the traceability mechanism provided by the process, which was based on human expert knowledge and the project background information in a similar fashion to the \emph{Arabidopsis} isoprenoids application example in Section~\ref{sec:isoprenoids}. The objectives measurability and their success criteria were set to the classification of 70\% of the samples correctly with acceptable true positive predictions across the predicted classes. This is the stage where it would be considered but for this exercise, there were no effective constraints. The dataset was found to be sufficient, adequate, and relevant to achieve the defined process objectives.

\begin{figure}[htb!h]
\centerline{\includegraphics[width=0.9\textwidth]{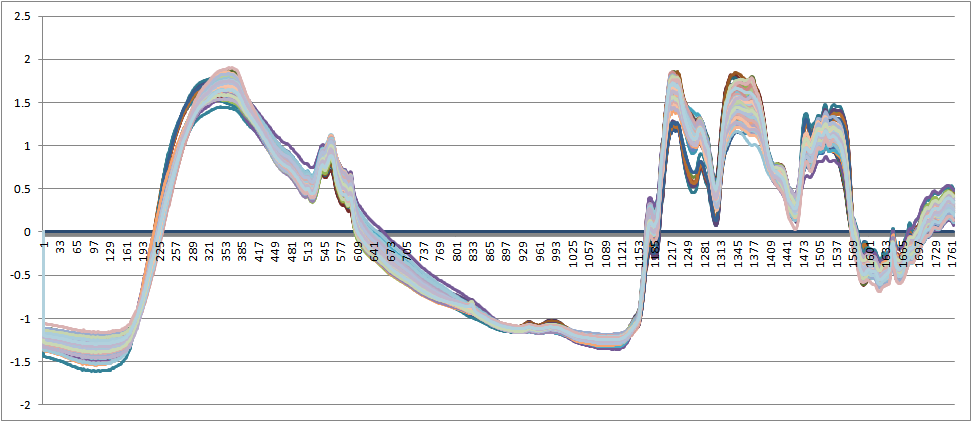}}
\centering
\caption{\textbf{Distribution of the Data in the \emph{Arabidopsis} Fingerprint Application:} A visualisation of the FT-IR metabolic fingerprint data.
\label{fig:HiMetFTIR_data_exploration_prospecting}}
\end{figure}

The dataset needed no extra \textbf{preprocessing}, as it was already preprocessed during data acquisition as described earlier. The \textbf{data exploration} phase was then executed. The data understanding uncovered that the number of the variables in the data is ten fold more than the number of the samples. The data prospecting showed that the fingerprint data visualisation is consistent with the conventional FT-IR spectrum as illustrated in Figure~\ref{fig:HiMetFTIR_data_exploration_prospecting}, while the data investigation showed no significant outliers or missing values.

\begin{figure}[htb!h]
\centerline{\includegraphics[width=0.9\textwidth]{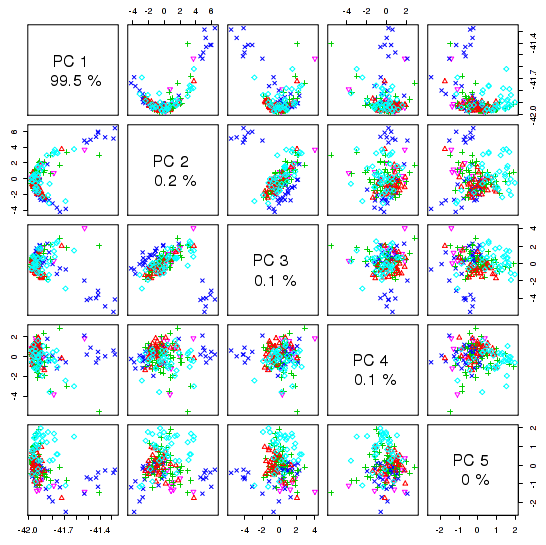}}
\centering
\caption{\textbf{Prospecting of the Data in the \emph{Arabidopsis} Fingerprint Application:} A score plot of 5 paired components and the proportion of variance which they explain.
\label{fig:HiMetFTIR_PCA}}
\end{figure}

On the other hand, PCA was used for prospecting the data. PCA is known for its ability to analyse high dimensional data and expressing it using a lower or reduced dimensionality. In this application, PCA was used for \textbf{data exploration} rather than modelling. It is used only for the purpose of prospecting the dataset attributes and exploring the reduced dataset potential for separating the samples. Figure~\ref{fig:HiMetFTIR_PCA} is showing the results of the data prospecting using PCA analysis, which uses score plots to illustrate the 5 most significant components showing their explained variance and paired distributions. PCA was able to separate the samples belong to some but not to all the prediction classes. This can be explained by the dominance of one component over the other as PC1 represents 99.5\% of the variance in the data.

The \textbf{technique selection} phase was executed in order to select a data mining technique that would have the potential of fulfilling the process classification objectives and is able to handle the massive number of variables using much fewer samples. PLS-LDA was selected and justified as potentially suitable for achieving the defined process objectives and handling the data high dimensionality. The requirements of the technique application were then assessed as well as the feasibility of its application.

No significant \textbf{acclimatisation} was needed apart from the limited data randomisation and splitting procedures. 66.0\% of the data was allocated for \textbf{model building} and training, while the remaining 34.0\% was allocated for testing the model. The \textbf{model building} customised phase was executed and the model performance was then recorded and reported by this phase.

\begin{figure}[htb!h]
\centerline{\includegraphics[width=0.5\textwidth]{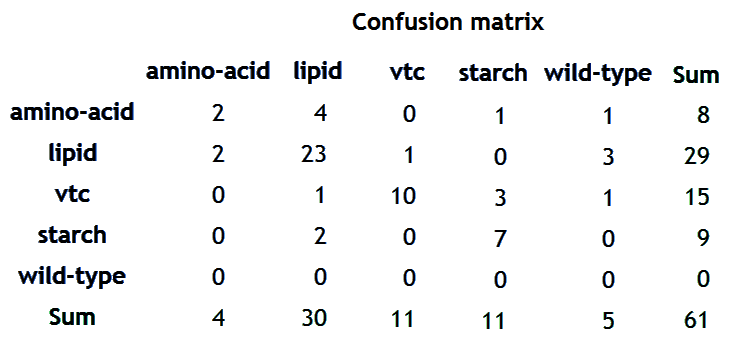}}
\centering
\caption{\textbf{Evaluation of the PLS-LDA Model in the Arabidpesis Fingerprint Application:} A confusion matrix showing the prediction accuracy of the PLS-LDA model.
\label{fig:HiMetFTIR_PLS-LDA_model_evaluation}}
\end{figure}

\begin{figure}[htb!h]
\centerline{\includegraphics[width=0.5\textwidth]{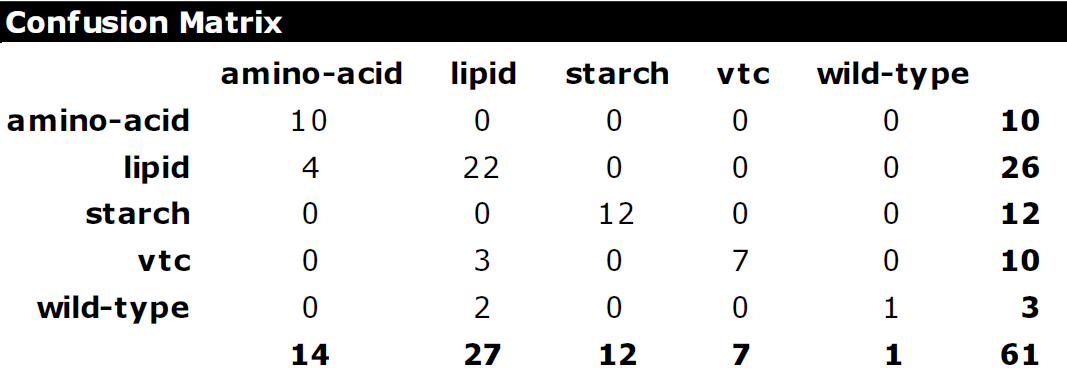}}
\centering
\caption{\textbf{Evaluation of the SOM Model in the \emph{Arabidopsis} Fingerprint Application:} A confusion matrix showing the prediction accuracy of the SOM model.
\label{fig:HiMetFTIR_som_model_evaluation}}
\end{figure}

The \textbf{model evaluation} phase was then executed and the performance of the model was then compared with measurability criteria, which was set in the \textbf{objectives definition} phase. The \textbf{model evaluation} had scored an error rate of 31\% and classification accuracy of 69.0\%. Figure~\ref{fig:HiMetFTIR_PLS-LDA_model_evaluation} illustrates the model confusion matrix. The model failed the success criteria of the process objectives, which was set to 70\% prediction accuracy. Therefore, feedback was performed from the \textbf{model evaluation} phase to \textbf{technique selection} phase in order to select an alternative technique. The \textbf{technique selection} phase was then iterated and its internal tasks were re-executed. Self-Organizing Maps (SOM) \cite{Ste06} was selected as a candidate modelling technique. After selecting the new technique, the \textbf{data acclimatisation} phase was revisited in order to ensure the suitability of the data to the new modelling technique. However, no significant change was made to the phase or to its generated data. The \textbf{model building} and \textbf{model evaluation} phases were re-executed. The SOM model outperformed the one built using PLS-LDA, with a classification accuracy of 85.0\% and with an error rate of less than 15.0\% and thus passed the evaluation criteria. The confusion matrix of the \textbf{model evaluation} is illustrated in Figure~\ref{fig:HiMetFTIR_som_model_evaluation}.

\begin{figure}[htb!h]
\centerline{\includegraphics[width=0.9\textwidth]{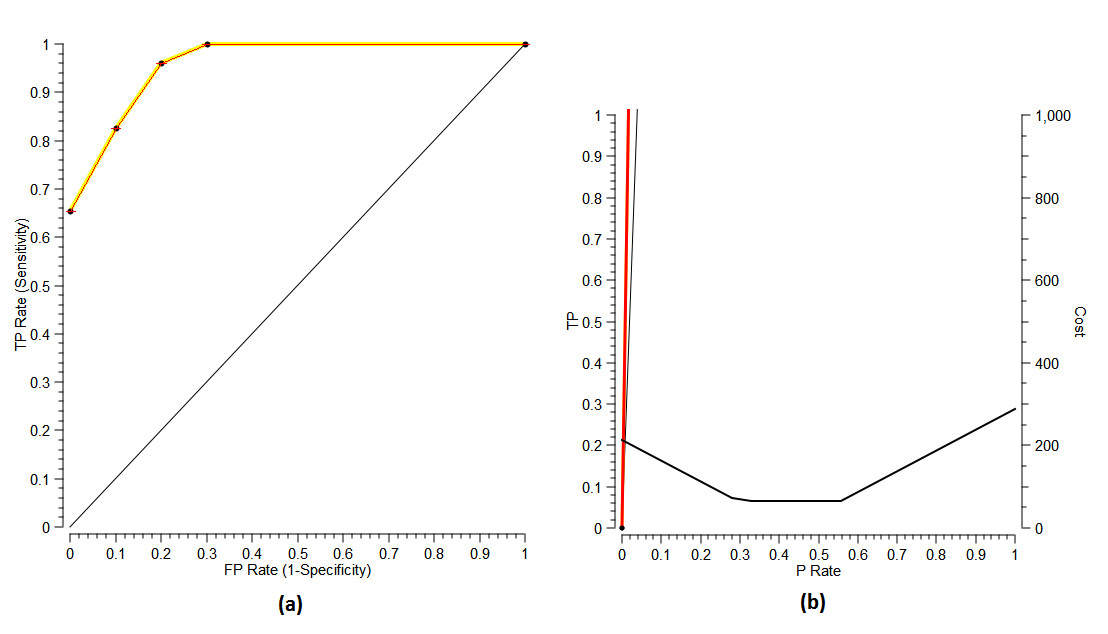}}
\centering
\caption{\textbf{The SOM \textbf{model evaluation} using ROC and Lift charts:} (a) ROC chart. The red curve represents the relationship between the true positive rate and false positive rate, and the yellow curve represents the ROC convex hull and provides an indication regarding the model sensitivity. (b) Lift chart. The red curve represents the relationship between true positive and positive values, while the black curve represents the cost function
\label{fig:HiMetFTIR_SOM_model_evaluation_ROC_Lift}}
\end{figure}

\begin{figure}[htb!h]
\centerline{\includegraphics[width=0.7\textwidth]{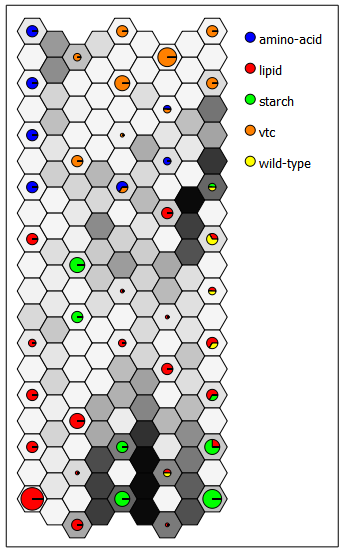}}
\centering
\caption{\textbf{Visualisation of the SOM Model in the \emph{Arabidopsis} Fingerprint Data:} A u-matrix map visualises the SOM model. The gray scale represents the distance between classes, where the darker colours represent larger distances. The colours in the cells represents the classes which their vectors belong to the cells, while the pie chart represents the ratio between the different classes which their vectors fall into the cell \cite{Ste11}.
\label{fig:HiMetFTIR_SOM_model_visualisation}}
\end{figure}

\textbf{Knowledge presentation} phase was then executed. The SOM model was presented and visualised using the U-matrix map, which is shown in Figure~\ref{fig:HiMetFTIR_SOM_model_visualisation}. The \textbf{knowledge evaluation} phase was then executed by performing the phase customised tasks, where the classification knowledge was found to be a fulfilment of the defined process objectives and was confirmed by the domain expert in terms of its validity and consistency with the metabolomics principles and background knowledge.

The results of the process were then deployed including the deliveries related to the SOM model, after executing the \textbf{deployment} phase and executing its tasks. The \textbf{deployment} of the results was based on the file system all deliveries were reported for publication in this article. The process execution was found valid in terms of its performance and compliance with MeKDDaM process model flow and structure and its practical aspects. The cross-delivery validation results were found positive. They were performed according to the mechanism defined by MeKDDaM process model.

\section{Kidney Disease Application}
This application involves investigating the difference between samples collected from patients with sever kidney disease and a healthy control group. It aims to analyse the difference between the two groups based on the metabolic fingerprint, which was acquired using NMR by researchers in Alberta University\cite{Psi07}.

\subsection{Materials and Methods}
\subsubsection{Experimental Protocol}
Before collecting the urine samples, the subjects were asked to fast the night before. They were also asked to avoid taking any medications, drinking alcohol or eating food such as fish, which is known to increase the level of metabolites, for the period of the 24 hours before the sample collection. The urine separation was performed through centrifugation at 1500g for a period of 15 minutes. The bacterial contamination of the sample was avoided by adding Sodium azide and storing the samples at a temperature of -80\textcelsius, which also helped to suppress the sample metabolism. More details regarding the study design are available in \cite{Psi07}.

\subsubsection{Data Acquisition}
The data acquisition was performed using a 500MHz NMR spectrometer, which was run at 300K. The NMR spectrum was then transformed using Fourier transformation after multiplying the FIDs by an exponential line broadening function with 0.3Hz, and after applying a sine-bell squared function. Phase and baseline correction was also applied to the acquired spectra manually by fitting the polynomial curve using the software instrument software \cite{Psi07}.

\subsection{Results of MeKDDaM Process Model Execution}
The process \textbf{objectives}  were defined to separate the data into two groups using the NMR metabolic fingerprint. The first corresponding to patients with severe kidney disease, while the second corresponding to the healthy group. These were defined in hypothesis-driven fashion, where a difference is assumed between the two groups, which can be noticeable in the NMR fingerprint data. The data was already \textbf{preprocessed} using the typical NMR procedures: baseline correction, filtering, Fourier transformation and peaks alignment.

\begin{figure}[htb!h]
\centerline{\includegraphics[width=0.7\textwidth]{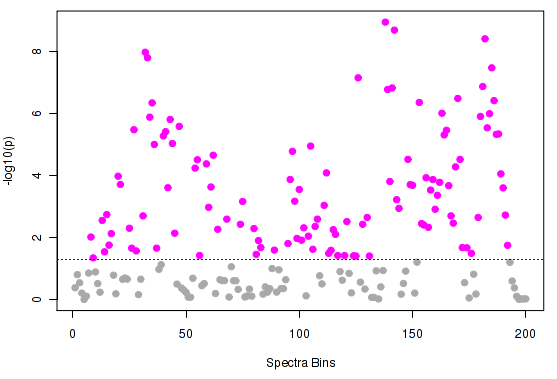}}
\centering
\caption{\textbf{Prospecting of the data in the Kidney Disease Application:} Hypothesis testing applied on the data using t-test. The results illustrates the difference based on a p-value threshold of 0.05 which is shown over the line. The p-values were transformed with -log10.
\label{fig:kidney_t-test}}
\end{figure}

\textbf{Data exploration} was carried out in order to investigate the quality of the data, validate its understanding, and prospect its potential. The data include 1.7\% zero values, while the data distribution was skewed to the left. The data were prospected regarding its potential towards differentiating the two sample groups using multiple univariate t-test. The results of data prospecting confirmed the data potential towards achieving the defined process objectives and provided insight regarding the difference between the two groups as illustrated in Figure~\ref{fig:kidney_t-test}.

Based on the \textbf{data exploration} report, SVM was found as a potential technique that would fulfil the process objectives and suits the nature of data. The data was then \textbf{acclimatised} based on the \textbf{data exploration} report to suit the requirements of the SVM technique. The zero values in the data were replaced by the mean value of each column as the data was found to have 1.7\% of zero values, while the distribution of the data was normalised using range scaling as the data distribution was found skewed to the left. Range scaling and auto-scaling were both recommended by \cite{Van06} as good data normalisation methods, who compared them to several other techniques, which was applied to the data in real-world metabolomics applications. The data were randomised and then divided into two halves, one was allocated for \textbf{model building} and training, while the other was reserved for \textbf{model evaluation}. The selection of this ratio was influenced by the small number of data instances, as a sufficient number of instances must be available for validating the model and ensure its ability to generalise. These decisions were recorded as part of the phase output.

\begin{figure}[htb!h]
\centerline{\includegraphics[width=0.5\textwidth]{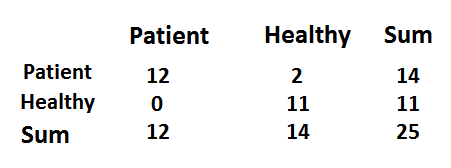}}
\centering
\caption{\textbf{Evaluation of the SVM Model in the Kidney Disease Application:} A confusion matrix showing the prediction accuracy of the SVM model.
\label{fig:kidney_svm_model_evaluation}}
\end{figure}

In the \textbf{model building} phase, the model was built based on the training dataset, which was acclimatised earlier. The model was exported to the file system as a PMML file as an example of standards compliance and then imported by the process implementation software \cite{Ban19a,Ban19c}. The \textbf{model evaluation} phase was then executed in order to validate the model based on the testing dataset. The model was successful in classifying 92\% of the examples with an error rate of 8\% which indeed passed the defined success criteria in the \textbf{objectives definition} phase. The confusion matrix of the \textbf{model evaluation} is illustrated in Figure~\ref{fig:kidney_svm_model_evaluation}.

The \textbf{knowledge presentation} phase. \textbf{Knowledge evaluation} was used in order to validate the model classification results in terms of their satisfaction of the process objectives and its consistency with the domain expert experience and the background knowledge. The \textbf{deployment} phase was then excused, where all the results were decided to deployed using relational database management systems as a deployment mechanism. However, due to the time constraints imposed by the project and since creating and maintaining a database system would take time, effort, and cost, the phase was iterated and the results were alternatively deployed to the file system for publication by this article. The \textbf{process evaluation} phase was then performed and the execution of the process was found valid and consistent with the defined process flow, structure and practical aspects as well as the cross-delivery validation results.

\section{Discussion}
The applications demonstrated MeKDDaM process model coverage of both data-driven and hypothesis-driven data mining approaches and its fulfillment of a number of analytical objectives. In the \emph{Arabidopsis} isoprenoids application, both C4.5 and MLP were used for predicting the samples classification, while in the cow diet application, random forest was used for describing the data important variables. In the kidney disease application, t-test was for verifying the the difference between the healthy and patient groups, which was then confirmed by the SVM classification.

\begin{figure}[htb!h]
\centerline{\includegraphics[width=0.6\textwidth]{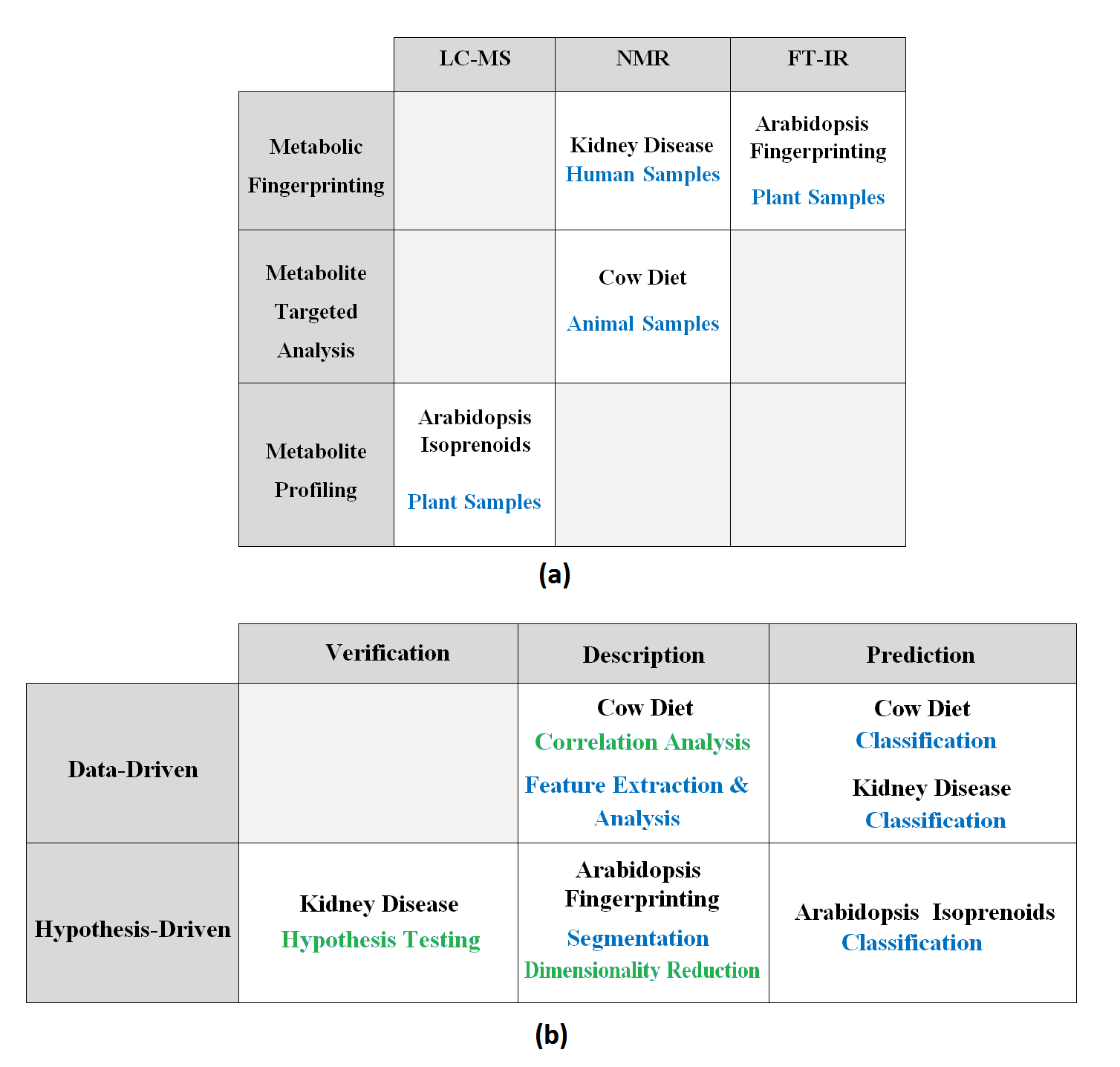}}
\centering
\caption{\textbf{The Applications Used for demonstrating the Process Execution:} (a) The applications' coverage of the metabolical approaches (rows) and the popular applied Data Mining Techniques (columns) (b) The applications' coverage of data mining approaches (rows), goals (columns) and tasks (cells). The tasks in blue were used for \textbf{model building}, while the ones in green were used as part of the exploratory analysis which was carried during data prospecting.
\label{fig:applications}}
\end{figure}

The applications also demonstrated the process model applicability using a number of data mining techniques and to perform a range of data mining tasks including classification, segmentation, hypothesis testing, correlation analysis, dimensionality reduction, and feature extraction and analysis. However, some of these tasks were applied only in order to perform exploratory data analysis as part of data prospecting such as correlation analysis, dimensionality reduction, hypothesis testing, while others were applied in the context of \textbf{model building} such as classification, segmentation, and feature extraction and analysis. Figure~\ref{fig:applications} summarises the applications discussed in this work and illustrates their coverage of metabolomics instruments and approaches as well as their coverage of data mining approaches, goals and tasks.

In addition, the applications demonstrated the process applicability to handle different metabolomics datasets that were acquired using the three major data acquisition instruments: LC-MS, FT-IR and NMR to handle various data pr-eprocessing and pretreatment requirements. The scientific orientation was demonstrated by applying the various traceability and justification mechanisms described by the process. The justification and traceability mechanisms provided by the software contributed to the consistency of the analysis results.

The process execution also confirmed the validity of the process flow and structure and the validity of the prescribed tasks within the process phases. It also demonstrated cohesion of the process phase internal tasks and understandability. The process also demonstrated the process adoption of the concept of mapping as well as the process of practical supplements and traceability.

The concept of mapping was also demonstrated through implementing a customisation mechanism for phase objectives, prerequisites and planning which are formulated based on the specific description and needs of each application. The customised activities were then loaded and populated in each phase before its execution. The implementation of this mechanism was found vital for improving the process efficiency and agility and for saving considerable time during the process execution. For example, the \emph{Arabidopsis} isoprenoids and the \emph{Arabidopsis} fingerprinting applications were applied on two different datasets which were acquired by different assays, but as part of the same metabolomics study. The process customisation in the former was reused during the execution of the latter application as they both shared the same study design, management resources, and sources of information. In addition, both applications also shared similar customisation in terms of the prerequisites, objectives and planning of their customised process phases. This saved more than one-third of the spent in executing the \emph{Arabidopsis} fingerprinting application. The software environment that was used during the process provided facilities for storing, exporting, and importing the phases customisation which was also found useful for the purpose of customising other aspects in the process such as supplement description and traceability mechanisms \cite{Ban12b,Ban19a}.

The process model iteration, feedback, and rollback mechanism were also practically useful. They helped to organise the forking of the process flow, while maintaining the validity and consistency of the execution of its phases and deliveries. Phase feedback and iteration were particularly useful for building alternative models as well as for updating the execution of the process phases, enhancing their performance or improving the quality of their deliveries. An example of the feedback during the process execution was described in the \emph{Arabidopsis} FT-IR application. The process iteration was also found useful for re-executing the process in order to answer a different question as demonstrated in the cow diet application. Yet, the rollback mechanisms, which was described by MeKDDaM process model and implemented by its realisation software was found useful. For example, in the \emph{Arabidopsis} isoprenoids application rollback was used for recovering the initial C4.5 model which outperformed the alternative MLP model.

The process execution demonstrated its support of management and planning on the level of the process and on the level of internal tasks within phases. The process execution demonstrated the realisation of the process resources management and allocation which was considered in almost every aspect of the process execution. As an example, during the demonstration of \emph{Arabidopsis} isoprenoids, the resources for the process phases were assigned and allocated for each planned activities. This covered the quantity, cost, type, and description of the software and hardware requirements as well as human expertise. Human interaction was also realised through assigning the performer of each planned activity during the phases' execution and through providing details regarding hi/her role, level of expertise, name and contact details. This was also useful for assigning human expertise as a source in the demonstration of traceability and justification mechanisms provided in the process execution. Examples of human interaction were demonstrated by the \emph{Arabidopsis} isoprenoids application as the performer of each activity in the process phases was pre-assigned and later recorder which was useful for demonstrating results and procedures traceability and justification as well as for demonstrating human interaction and manageability and resources manageability. In addition, the execution of the process demonstrated the various quality assurance mechanisms provided by the process and its implementation software in order to satisfy the requirements defined. The quality of the data is investigated in the \textbf{data exploration} phase and handled later in \textbf{data acclimatisation} as demonstrated in the kidney disease application. In addition, the validity of the model is evaluated in the \textbf{model evaluation} phase as demonstrated in all the four applications.

Metabolomics and data mining standards were both considered in the development of MeKDDaM process model as well as in the deliveries of the process phases. This was demonstrated by the application through the performance of both \textbf{data pre-processing} and pretreatment procedures, which have been considered in the designing of the \textbf{data pre-processing} and \textbf{data acclimatisation} phases as well as in their realisation by a software environment \cite{Ban19a,Ban19c}. However, the support for reporting standards e.g. PMML depended on the Data Management Group (DMG) support of the particular \textbf{model building} technique \cite{Dmg10} as well as on the ability of external data mining tool for delivering the model in a particular standardized format. In the kidney disease application, the SVM model was successfully delivered in PMML format. However, the application of the adopted standards is still constrained by the facilities provided by the \textbf{model building} tool. In this case, the model can still be coded in a PMML format independently as long as the technique is supported by DMG.

The visualisation was particularly useful in presenting the C4.5 model built in the isoprenoids profiling application which was built with an informative tree-like representation as well as in the \emph{Arabidopsis} fingerprinting application, where the SOM model was presented using a U-matrix representation. In addition, visualisation was found as a powerful tool for prospecting the data. Examples of data visualisation were demonstrated in the cow diet application as clustering trees were used for prospecting the data as illustrated in Figure~\ref{fig:cow_raw_data_correlation} as well as in the \emph{Arabidopsis} isoprenoids application, where correlation heat map was used for prospecting the data as illustrated in Figure~\ref{fig:HiMet_data_exploration_prospecting}. In addition, the \emph{Arabidopsis} isoprenoids prospecting uncovered a significant correlation between some of the metabolites which were found later involved in carotenoid biosynthesis metabolic pathway during \textbf{knowledge evaluation}. The decision tree in Figure~\ref{fig:HiMet_C45_model_visualisation} provides an understandable presentation of the C4.5 model which can be interpreted by the domain/biologist even if he/she has limited knowledge in machine learning. The diagram also provided information regarding the relationship between the profile metabolites and provided useful statistical information regarding the confidence of the prediction in each class. The \emph{Arabidopsis} isoprenoid application demonstrated an example, where the classification results were used to put the knowledge in a metabolomics context as the metabolic pathway of the significant metabolites was investigated and found related the carotenoids biosynthesis. In the \emph{Arabidopsis} fingerprinting application, the U-matrix in Figure~\ref{fig:HiMetFTIR_SOM_model_visualisation} was successful in illustrating the clusters of the predicted classes as well as the distance between them. Furthermore, the applications reported in this work demonstrated the realisation of the process model \cite{Ban19a,Ban19c} which satisfies its automation requirements.

In addition, the applications demonstrated in this work performed all MeKDDaM process model execution scenarios illustrated by Figure~\ref{fig:ProcessExecutionScenariosCopy}. Scenario 1, 2, 3, and 4 was demonstrated in the \emph{Arabidopsis} isoprenoids application. Scenario 1 demonstrated an example of the process model feedback mechanism as the \textbf{data pre-processing} phase was iterated in order to perform further preprocessing procedures based on the results of \textbf{data exploration} phase. Scenario 2 demonstrates another feedback, which was performed in order to build an alternative model, while scenario 3 demonstrated the process rollback mechanism, where the original model was recovered as it outperformed the alternative model. In addition, scenario 5 was demonstrated in the kidney disease application, where an alternative deployment mechanism was chosen due to the project time limitation. Scenario 6 was performed in the cow diet application, where the process was iterated in order to answer a new question and to achieve different objectives  than those defined in the first iteration, while scenario 7 was performed in the \emph{Arabidopsis} application, where the process was instantiated and applied to be applied in the same metabolomics study as the \emph{Arabidopsis} isoprenoids application, but with a different dataset.

\section{Conclusion} \label{sec:conclsion}
The process applications demonstrated the process applicability to both data-driven and hypothesis-driven data mining approaches and its ability to achieve objectives defined based on various data mining goals including prediction, description and verification. The applications also demonstrated MeKDDaM process model validity to apply a number of data mining techniques to perform various data mining tasks including classification, segmentation, hypothesis testing, correlation analysis, dimensionality reduction, and feature extraction and analysis. The process applications demonstrate the process model applicability to a variety of metabolomics datasets acquired using three of the popular chemical analysis instruments: LC-MS, NMR, and FT-IR, which have been discussed earlier. The applications also provided coverage of the three major metabolic approaches: fingerprinting, metabolite profiling, and targeted analysis which was applied in a variety of metabolomics investigations which analysed plant, animal, and human materials. In addition, the applications demonstrated the process model ability to perform the various execution scenarios that were illustrated in Figure~\ref{fig:ProcessExecutionScenariosCopy}.

The process execution demonstrated the process model's support to data handling, preprocessing and pre-treatment. MeKDDaM satisfied the scientific orientation requirements through its support to various traceability and justification mechanisms. The process execution confirmed the validity of the process flow and structure and the validity of the prescribed tasks within the process phases. The process phases and their internal tasks were found to be both cohesive and understandable, thanks to the process concept of mapping, where the phases generic tasks were customized to suit the needs of the application as well as the process practical supplements and traceability. The applications provided examples of the process supplements and practical features: management, human interaction, quality assurance and standards.

The applications demonstrate the process support for resources allocation and for the management of human interaction, such as assigning the performer of process activities and assigning domain experts who act as a source in the traceability and justification mechanisms. The process execution demonstrated support for the various aspects of quality assurance regarding the process data, procedures and deliveries. The quality of the data is investigated in the \textbf{data exploration} phase as well as in \textbf{data acclimatisation} phase. The validity of the model is evaluated in the \textbf{model evaluation} phase. The opportunities for complying with metabolomics and data mining standards are illustrated. The deliveries of the process phases were designed to comply with the MSI reporting standards. The applications demonstrate the process support for visualisation as the preferred \textbf{knowledge presentation} mechanism.

The process execution also demonstrated the importance of the \textbf{data exploration} phase, which was vital for \textbf{technique selection} and \textbf{data acclimatisation}. The applications demonstrated the process model satisfaction of the requirement of \textbf{data exploration} and the software implementation of the various mechanisms used for its realisation such as data understanding, data investigation and data prospecting. The applications highlighted the process model's satisfaction of \textbf{knowledge presentation} requirements and reflected its importance for putting the knowledge into the context of metabolomics particularly in the two demonstrated \emph{Arabidopsis} applications.

\bibliographystyle{unsrt}

\end{document}

%% file: himet_c45_model_visualisation.tex
\begin{landscape}
\pagestyle {plain}
\scriptsize

\begin{figure}[tb!h]
\centerline{\includegraphics[width=1.45\textwidth]{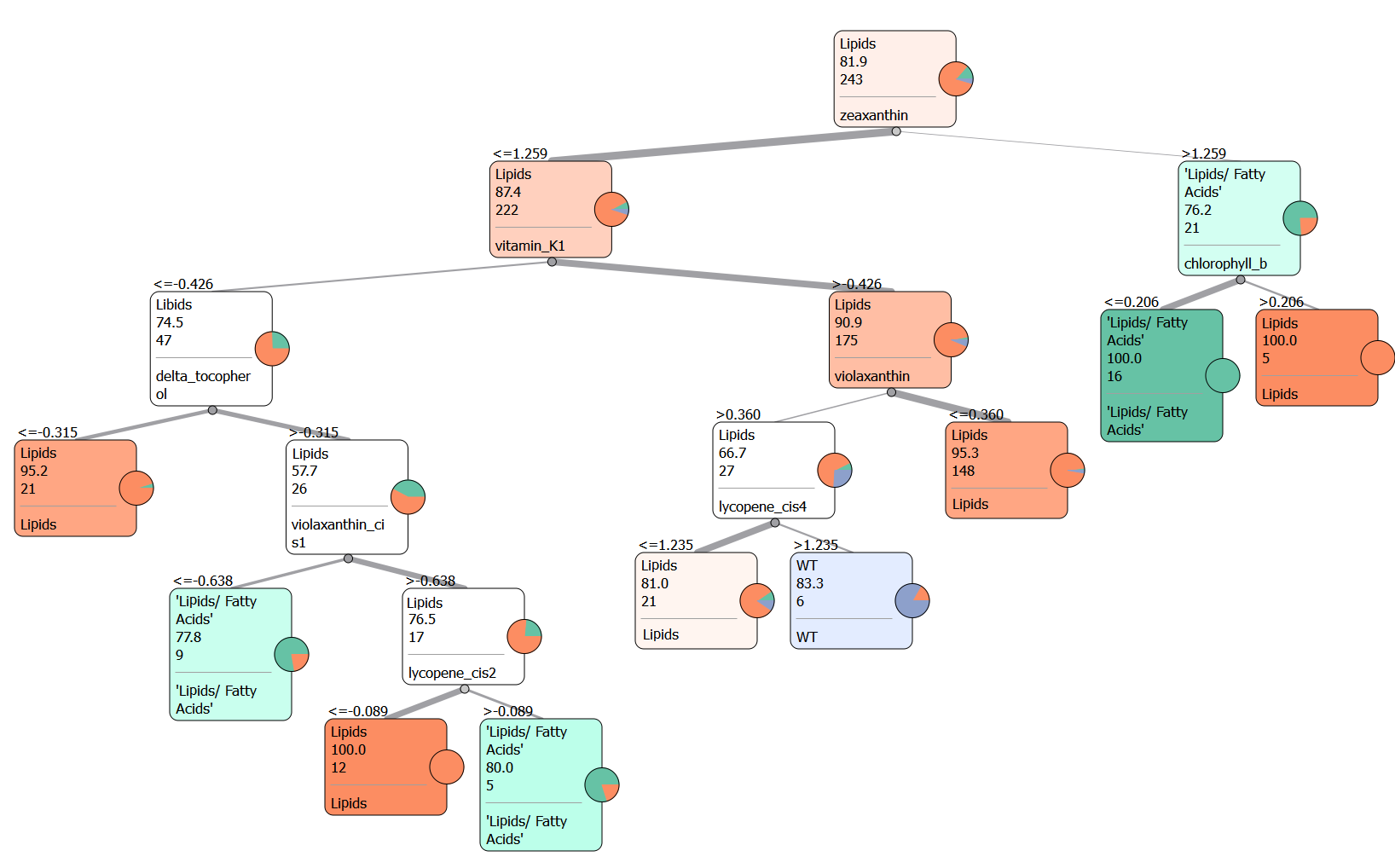}}
\centering
\caption{\textbf{Visualisation of the C4.5 model in the \emph{Arabedopesis Isoprenoids Application}:} An illustration of the C4.5 decision tree model
\label{fig:HiMet_C45_model_visualisation}}
\end{figure}

\end{landscape}